\pdfoutput=1
\documentclass[reprint, aps, prd, preprintnumbers, amsmath, amssymb, superscriptaddress, nofootinbib]{revtex4-1}

\usepackage{amssymb,amsmath,bbold,slashed}
\usepackage{graphicx,soul}
\usepackage[usenames,dvipsnames]{xcolor}
\usepackage[unicode=true,pdfusetitle,
 bookmarks=true,bookmarksnumbered=false,bookmarksopen=false,citecolor=Turquoise,
 breaklinks=false,pdfborder={0 0 1},backref=false,colorlinks=true,pdfpagemode=FullScreen]
 {hyperref}

\usepackage{tikz}

\usetikzlibrary{arrows,calc,graphs,patterns,positioning}
\usetikzlibrary{decorations,decorations.markings,decorations.pathmorphing,decorations.pathreplacing}
\usetikzlibrary{shapes, shapes.geometric}

\tikzset{/pgf/decoration/.cd,
    number of sines/.initial=5,
    angle step/.initial=20,
}
\newdimen\tmpdimen

\pgfdeclaredecoration{complete sines}{initial}
{
    \state{initial}[
        width=+0pt,
        next state=move,
        persistent precomputation={
            \pgfmathparse{\pgfkeysvalueof{/pgf/decoration/angle step}}%
            \let\anglestep=\pgfmathresult%
            \let\currentangle=\pgfmathresult%
            \pgfmathsetlengthmacro{\pointsperanglestep}%
                {(\pgfdecoratedremainingdistance/\pgfkeysvalueof{/pgf/decoration/number of sines})/360*\anglestep}%
        }] {}
    \state{move}[width=+\pointsperanglestep, next state=draw]{
        \pgfpathmoveto{\pgfpointorigin}
    }
    \state{draw}[width=+\pointsperanglestep, switch if less than=1.25*\pointsperanglestep to final, 
        persistent postcomputation={
        \pgfmathparse{mod(\currentangle+\anglestep, 360)}%
        \let\currentangle=\pgfmathresult%
    }]{%
        \pgfmathsin{+\currentangle}%
        \tmpdimen=\pgfdecorationsegmentamplitude%
        \tmpdimen=\pgfmathresult\tmpdimen%
        \divide\tmpdimen by2\relax%
        \pgfpathlineto{\pgfqpoint{0pt}{\tmpdimen}}%
    }
    \state{final}{
        \ifdim\pgfdecoratedremainingdistance>0pt\relax
            \pgfpathlineto{\pgfpointdecoratedpathlast}
        \fi
   }
}

\tikzset{
    scalar/.style={draw=black, thick, dashed},
    fermion/.style={draw=black, postaction={decorate}, decoration={markings, mark=at position .55 with {\arrow[]{latex}}}, thick},
    antifermion/.style={draw=black, postaction={decorate}, decoration={markings, mark=at position .55 with {\arrow{latex}}}, thick},
    majorana/.style={draw=black, thick},
    photon/.style={draw=none, outer ysep=3pt, postaction={draw=black, thick, decorate}, decoration={complete sines, amplitude=7pt}},
    photonloop/.style={draw=none, outer ysep=3pt, postaction={draw=black, thick, decorate}, decoration={complete sines, number of sines=15, amplitude=5pt}},
    gluon/.style={draw=none, outer ysep=3pt, postaction={draw=black, thick, decorate}, decoration={coil, amplitude=3pt, segment length=5pt}}, 
    composite/.style={draw=gray!70!, line width=5pt},
    vtx/.style={inner sep=0pt},
    mass/.style={cross out, draw, minimum size=5pt, inner sep=0pt},
    ins/.style={draw, cross, shape=circle, minimum size=5pt, inner sep=0pt}, 
    cross/.style={path picture={\draw[black] (path picture bounding box.south east) -- (path picture bounding box.north west) (path picture bounding box.south west) -- (path picture bounding box.north east);}},
    vertex/.style={draw, shape=circle, fill=black, minimum size=3pt, inner sep=0pt},
    blob/.style={draw, shape=circle, preaction={fill,white}, pattern = north west lines, minimum size=15pt, inner sep=0pt},        
    loop/.style={shape=circle, minimum size=1.7cm, outer sep=9pt},
    site/.style={draw, shape=circle, minimum size=1.5cm, inner sep=0pt},
    brane/.style={trapezium, draw, trapezium stretches=true, minimum height=4cm, minimum width=3.5cm, inner sep=0, trapezium left angle=35, trapezium right angle=145, shape border uses incircle, shape border rotate=17.5},
    momentum/.style={->, dist/.store in=\segDistance, pos/.store in=\segPos, len/.store in=\segLength,
      to path={
      ($(\tikztostart)!\segPos!(\tikztotarget)!\segLength/2!(\tikztostart)!\segDistance!90:(\tikztotarget)$) -- 
      ($(\tikztostart)!\segPos!(\tikztotarget)!\segLength/2!(\tikztotarget)!\segDistance!-90:(\tikztostart)$)  \tikztonodes
      }, 
      pos=.5,
      len=7mm,
      dist=2mm
    },    
}

\begin{document}

\preprint{IPMU17-0138}

\title{Right-handed Neutrino Dark Matter in a $U(1)$ Extension of the Standard Model}

\author{Peter Cox}
\email[]{peter.cox@ipmu.jp}
\affiliation{Kavli IPMU (WPI), UTIAS, University of Tokyo, Kashiwa, Chiba 277-8583, Japan}
\author{Chengcheng Han}
\email[]{chengcheng.han@ipmu.jp }
\affiliation{Kavli IPMU (WPI), UTIAS, University of Tokyo, Kashiwa, Chiba 277-8583, Japan}
\author{Tsutomu T. Yanagida}
\email[]{tsutomu.tyanagida@ipmu.jp}
\affiliation{Kavli IPMU (WPI), UTIAS, University of Tokyo, Kashiwa, Chiba 277-8583, Japan}
\affiliation{Hamamatsu Professor}

\begin{abstract}
We consider minimal $U(1)$ extensions of the Standard Model in which one of the right-handed neutrinos is charged under the new gauge symmetry and plays the role of dark matter. 
In particular, we perform a detailed phenomenological study for the case of a $U(1)_{(B-L)_3}$ flavoured $B-L$ symmetry. 
If perturbativity is required up to high-scales, we find an upper bound on the dark matter mass of $m_\chi\lesssim2$\,TeV, significantly stronger than that obtained in simplified models. 
Furthermore, if the $U(1)_{(B-L)_3}$ breaking scalar has significant mixing with the SM Higgs, there are already strong constraints from direct detection. 
On the other hand, there remains significant viable parameter space in the case of small mixing, which may be probed in the future via LHC $Z^\prime$ searches and indirect detection.
We also comment on more general anomaly-free symmetries consistent with a TeV-scale RH neutrino dark matter candidate, and show that if two heavy RH neutrinos for leptogenesis are also required, one is naturally led to a single-parameter class of $U(1)$ symmetries.
\end{abstract}

\maketitle

\section{Introduction}

The Standard Model (SM) with the addition of heavy Majorana right-handed (RH) neutrinos provides a very successful model for explaining low-energy observations. 
As is well-known, small neutrino masses are naturally generated via the seesaw mechanism \cite{Minkowski:1977sc, *Yanagida:1979as, *Glashow:1979nm, *GellMann:1980vs}, and the observed baryon asymmetry is dynamically created through leptogenesis in the early universe \cite{Fukugita:1986hr}. 
The last remaining low-energy observation demanding the presence of new physics is the existence of dark matter (DM); if this minimal extension of the SM could also provide a viable DM candidate, it would therefore present a very attractive possibility. 

This is in fact the case in the $\nu MSM$~\cite{hep-ph/0505013}, where the lightest RH neutrino has a keV-scale mass and is the dark matter~\cite{hep-ph/9303287}. 
However, in this scenario one is forced to pay a relatively high price; the benefit of the seesaw mechanism is largely lost, and one must resort to resonant leptogenesis with highly degenerate masses for the two GeV-scale RH neutrinos. 
In this work, we present an alternative scenario where the lightest RH neutrino plays the role of DM, while retaining the standard high-scale seesaw and thermal leptogenesis. 
In the cases we consider, this is achieved via the introduction of a new $U(1)$ gauge symmetry that is spontaneously broken at low energies; however, there also exist other interesting possibilities~\cite{1006.1731}. 

Given that only two superheavy ($M_{\nu_R}\gtrsim10^{9}\,$GeV) RH neutrinos are required to generate the observed light neutrino masses and participate in leptogenesis~\cite{hep-ph/0208157}, there is no fundamental barrier to assuming that the third RH neutrino ($\nu_R^3$) is significantly lighter and a dark matter candidate; indeed, one could argue that the existence of DM provides the only reason to include the third RH neutrino at all~\cite{1602.03003}. 
However, there remain several issues that must be addressed: (i) stability of the DM; (ii) $\nu_R^3$ production in the early universe; and (iii) the large hierarchy in the RH neutrino masses. 
The first issue is closely related to two other problems: the tendency of $\nu_R^3$ to washout the previously generated lepton asymmetry, and to significantly raise the scale of the light active neutrino masses. 
These problems can be avoided, and DM stability ensured, by simply imposing a $\mathbb{Z}_2$ symmetry where $\nu_R^3$ is the only field with negative parity. 
The second and third issues above can be simultaneously addressed by introducing a new $U(1)$ gauge symmetry under which $\nu_R^3$ is charged. 
A large Majorana mass for $\nu_R^3$ is then forbidden by the gauge symmetry and is only generated upon spontaneous breaking of the $U(1)$, while the new gauge interactions provide a production mechanism for $\nu_R^3$ in the early universe. If the $U(1)$ breaking scale is $\sim\,$TeV, then the $\nu_R^3$ mass is naturally of the same order, and it can be produced via thermal freeze-out. In other words, the $\nu_R^3$ becomes a natural WIMP dark matter candidate. 

In this paper, we begin by discussing possible choices for the $U(1)$ gauge symmetry that are consistent with the above picture. 
We then focus on one particularly interesting model, a flavoured $B-L$ symmetry: $U(1)_{(B-L)_3}$~\cite{1704.08158, 1705.00915, 1705.01822, 1705.03858}. 
This model also represents the least constrained model within our framework and as such provides a good case for a detailed phenomenological study. 
In Sections~\ref{sec:constraints} \& \ref{sec:results}, we discuss in detail the current theoretical and experimental constraints on the parameter space of this model, and comment briefly on the future experimental prospects. 
Our main results are then summarised in Sec.~\ref{sec:conclusion}.

\section{$U(1)$ extensions within $SM+3\nu_R$}

We are interested in $U(1)$ extensions of the SM that can provide a thermal production mechanism for $\nu_R^3$ in the early universe, while still allowing two RH neutrinos to have large Majorana masses.\footnote{If one removes the assumption of high-scale leptogenesis there are many additional possibilities, for example the usual $U(1)_{B-L}$~\cite{1002.2525, 1601.07526, 1606.09317, 1611.02672}.} 
Naively, one might think that there are many such gauge symmetries; however, as we shall now demonstrate, the combination of anomaly cancellation (we assume there are no additional chiral fermions charged under the SM gauge group) and phenomenological considerations is in fact quite restrictive.

For simplicity, we shall restrict ourselves to considering vectorial symmetries (i.e. LH and RH fields have the same charge, modulo $U(1)_Y$). 
However, such an assumption is also reasonably well-motivated, since for chiral symmetries where the SM Higgs is charged under the $U(1)$, $Z$-$Z^\prime$ mass mixing occurs at tree-level and can lead to strong constraints from electroweak precision measurements. 
This single additional assumption, combined with the requirement that exactly two RH neutrinos can have large Majorana masses and give rise to leptogenesis, immediately fixes the charges (up to overall normalisation) in the lepton sector. 
In flavour space we have\footnote{The case $Q_l = (x,-x,-1)$ also allows for two heavy RH neutrinos; however, they form a Dirac pair and do not generate a lepton asymmetry.}
\begin{equation} \label{eq:Q_l}
  Q_l = (0,0,-1) \,.
\end{equation}

Moving now to the quark sector, one needs to be careful about possible flavour-changing neutral currents (FCNC) mediated by the $Z^\prime$, in particular due to the very stringent constraints from $K-\bar{K}$ mixing. 
A natural way to evade these constraints is to impose universal charges for the first and second generation quarks. 
Anomaly cancellation then fixes the charge of the third generation quarks in terms of the first two . Explicitly, we find that
\begin{equation} \label{eq:Q_q}
  Q_q = (a, a, \frac{1}{3}-2a) \,.
\end{equation}

In principle, one could now perform a general phenomenological study of the above class of $U(1)$ symmetries. 
For generic values of $a$, existing two-mediator simplified models~\cite{1606.07609, 1605.09382} will be broadly applicable; however, there will be additional constraints arising from the non-zero charge of the leptons in Eq.~\eqref{eq:Q_l}. 
Instead, we will consider in detail the special case $a=0$, which is of particular interest for a number of reasons. 
Firstly, one should immediately notice that it in fact corresponds to a \emph{flavoured} $B-L$ symmetry, under which only a single generation of fermions is charged. 
It therefore preserves the nice feature of the SM that anomaly cancellation is satisfied independently within each generation.
In fact, one could consider it as the low-energy remnant of a $(U(1)_{{(B-L})_i})^3$ symmetry at high scales; such a symmetry could arise in the context of $[SO(10)]^3$ grand unification~\cite{0709.3491}. 
Interestingly, this model can also provide an explanation~\cite{1704.08158, 1705.03858, 1705.00915} of the recent hints of lepton universality violation in rare $B$ decays observed by LHCb~\cite{1406.6482, 1705.05802}. 
Lastly, it is also likely to be the least constrained model within our framework. This is due to the absence of couplings to the first and second generation quarks, which can significantly alleviate the bounds from LHC searches, and to a lesser extent direct detection. 
As we shall see, even this best-case scenario is already becoming increasingly constrained by experimental searches, although significant regions of parameter space still remain to be explored.

\section{Review of the Flavoured B-L Model}

We consider a $U(1)_{(B-L)_3}$ local symmetry under which only the third generation fermions are charged. 
The SM Higgs, $H$, is taken to be neutral under the new symmetry. 
We also introduce a new complex scalar, $\Phi$, with $U(1)_{(B-L)_3}$ charge $+2$. 
This field is responsible for spontaneously breaking $U(1)_{(B-L)_3}$, and simultaneously generating a Majorana mass for $\nu_R^3$. 
Finally, we impose a $\mathbb{Z}_2$ discrete symmetry under which $\nu_R^3$ takes odd parity, while all other fields are even. 
This forbids Yukawa couplings between $\nu_R^3$ and the SM fermions, and guarantees DM stability. 

\bgroup
\def\arraystretch{1.5}
\begin{table}[ht]
  \begin{tabular}{| c | c | c | c | c |}
    \hline
    & $q_L^3$, $b_R$, $t_R$ & $\ell_L^3$, $\tau_R$, $\nu_R^3$ & \hspace{0.25cm}$H$\hspace{0.25cm} & \hspace{0.25cm}$\Phi$\hspace{0.25cm} \\
    \hline
    $Q_{(B-L)_3}$ & +1/3 & -1 & 0 & +2 \\
    \hline
  \end{tabular}
  \caption{Charges under $U(1)_{(B-L)_3}$.}
  \label{tab:charges}
\end{table}
\egroup

Note that the above charge assignments clearly forbid Yukawa couplings between the third generation fermions and the first two generations. 
These couplings can be generated upon spontaneous symmetry breaking by either introducing additional Higgs doublets carrying $U(1)_{(B-L)_3}$ charges, or $U(1)_{(B-L)_3}$-neutral vector-like fermions that mix with the SM quarks and leptons; the latter possibility is described in detail in Ref.~\cite{1705.03858}. 
The non-trivial flavour structure of the $U(1)_{(B-L)_3}$ charges also leads to new physical mixing angles involving the third generation, in addition to those present in the CKM and PMNS matrices. 
Here, we simply assume that these angles are small and neglect them in our analysis. 
See Refs.~\cite{1705.03858, 1705.01822, 1705.00915} for a more detailed discussion, including the bounds from flavour observables. 

The details of the above mechanism can be largely neglected when discussing the DM phenomenology. 
The only assumption we make is that $\Phi$ dominates the $U(1)_{(B-L)_3}$ symmetry breaking (i.e. $\langle\Phi\rangle\gg$ vev of any additional scalars), so that we can make use of tree-level relations to reduce the number of free parameters. 

In summary, our model contains a new $Z^\prime$ gauge boson, a $U(1)_{(B-L)_3}$ breaking scalar $\Phi$, and a Majorana fermion dark matter candidate. 
Here, we define the four-component spinor $\chi=(-\varepsilon\nu^{3*}_R,\, \nu^3_R)^T$ to describe the Majorana dark matter. 
The relevant Lagrangian can be summarised as
\begin{align} \label{eq:lag}
  \mathcal{L}&= \frac{i}{2} \bar{\chi} \slashed{\partial} \chi+\frac{g}{2} Z^\prime_\mu \bar{\chi} \gamma^5 \gamma^\mu \chi- (\frac{y}{2} \bar{\chi} \Phi P_R\chi +h.c.)  \nonumber \\
                    &+ (D^\mu \Phi)^\dagger (D_\mu \Phi)+\mu^2_\Phi \Phi^\dagger \Phi-\lambda_\Phi(\Phi^\dagger \Phi)^2-\lambda_{H \Phi} (H^\dagger H)(\Phi^\dagger \Phi) \nonumber \\
                    &- \frac{1}{4} F^{\prime \mu \nu} F^\prime_{\mu \nu}- \frac{\epsilon}{2} F^{\prime \mu \nu} B_{\mu \nu} + \mathcal{L}_{SM} \,,
\end{align}
where $P_R$ is the RH projection operator. The two scalars acquire vacuum expectation values, such that in unitary gauge:
\begin{equation}
  \Phi= \frac{1}{\sqrt{2}}(w+\phi)\,, \quad H=\frac{1}{\sqrt{2}}(0,\, v+h)^T \,.
\end{equation}
In the limit $\lambda_{H \Phi} \rightarrow 0$, we then have
\begin{align} \label{eq:masses}
  m_\chi &= \frac{y}{\sqrt{2}}w \,, \notag \\
  m_{Z^\prime} &= 2 g w \,, \\
  m_\phi &= \sqrt{2} \mu_\Phi \notag \,, 
\end{align}
where  $w= \mu_\Phi/\sqrt{\lambda_\Phi}$. 
These expressions can be used to express the above Lagrangian in terms of the five free parameters $(g, m_\chi, m_{Z^\prime}, m_\phi, \epsilon)$. 
In particular, the DM Yukawa coupling can be expressed in terms of the gauge coupling as
\begin{equation} \label{eq:yukawa}
  y = \frac{2\sqrt{2} g m_\chi}{m_{Z^\prime}} \,.
\end{equation}

Lastly, non-zero values of $\lambda_{H \Phi}$ will introduce mixing between $\phi$ and $h$. 
This introduces one additional free parameter, namely the mixing angle $\theta$.
The details are given in Appendix~\ref{app:mixing}.

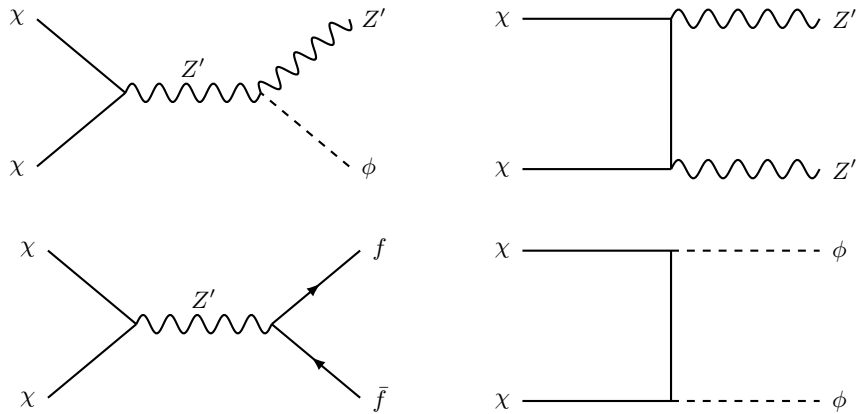
\begin{figure*}[t]
\centering
\begin{tikzpicture}
  \node[vtx, label=180:$\chi$] (i1) at (0,1) {};
  \node[vtx, label=180:$\chi$] (i2) at (0,-1) {};
  \coordinate[] (v1) at (1.2,0) {};
  \coordinate[] (v2) at (3.0,0) {};
  \node[vtx, label=0:$Z^\prime$] (o1) at (4.2,1) {};
  \node[vtx, label=0:$\phi$] (o2) at (4.2,-1) {};
  \graph[use existing nodes]{
     i1 --[majorana] v1 --[majorana] i2;
     o2 --[scalar] v2 --[photon] o1;
     v1 --[photon, edge label=$Z^\prime$] v2;
  };
\end{tikzpicture}
\hspace{1cm}
\begin{tikzpicture}
  \node[vtx, label=180:$\chi$] (i1) at (0,1) {};
  \node[vtx, label=180:$\chi$] (i2) at (0,-1) {};
  \coordinate[] (v1) at (2,1) {};
  \coordinate[] (v2) at (2,-1) {};
  \node[vtx, label=0:$Z^\prime$] (o1) at (4,1) {};
  \node[vtx, label=0:$Z^\prime$] (o2) at (4,-1) {};
  \graph[use existing nodes]{
     i1 --[majorana] v1 --[majorana] v2 --[majorana] i2;
     v1 --[photon] o1;
     v2 --[photon] o2;
  };
\end{tikzpicture}

\vspace{4ex}

\begin{tikzpicture}
  \node[vtx, label=180:$\chi$] (i1) at (0,1) {};
  \node[vtx, label=180:$\chi$] (i2) at (0,-1) {};
  \coordinate[] (v1) at (1.2,0) {};
  \coordinate[] (v2) at (3.0,0) {};
  \node[vtx, label=0:$f$] (o1) at (4.2,1) {};
  \node[vtx, label=0:$\bar{f}$] (o2) at (4.2,-1) {};
  \graph[use existing nodes]{
     i1 --[majorana] v1 --[majorana] i2;
     o2 --[fermion] v2 --[fermion] o1;
     v1 --[photon, edge label=$Z^\prime$] v2;
  };
\end{tikzpicture}
\hspace{1cm}
\begin{tikzpicture}
  \node[vtx, label=180:$\chi$] (i1) at (0,1) {};
  \node[vtx, label=180:$\chi$] (i2) at (0,-1) {};
  \coordinate[] (v1) at (2,1) {};
  \coordinate[] (v2) at (2,-1) {};
  \node[vtx, label=0:$\phi$] (o1) at (4,1) {};
  \node[vtx, label=0:$\phi$] (o2) at (4,-1) {};
  \graph[use existing nodes]{
     i1 --[majorana] v1 --[majorana] v2 --[majorana] i2;
     v1 --[scalar] o1;
     v2 --[scalar] o2;
  };
\end{tikzpicture}
\caption{Representative diagrams for the main DM annihilation channels.}
\label{fig:annihilation}
\end{figure*}

\section{DM Annihilation and Relic Density} \label{sec:relic_density}

Dark matter annihilation proceeds via the new $U(1)_{(B-L)_3}$ gauge interactions and the Yukawa interaction between $\chi$ and $\phi$. 
Representative diagrams for the four main annihilation channels are shown in Fig.~\ref{fig:annihilation}. 
Once $h$-$\phi$ mixing is included, there are the additional annihilation channels, $\chi\chi\to h\phi$, $\chi\chi\to hh$, and $\chi\chi\to\,$SM~SM  via an s-channel scalar. 
  
The expressions for the annihilation cross-sections in the limit $\theta\to0$ are given in Eqs.~(\ref{eq:xsec1}--\ref{eq:xsec4}). 
We have performed the usual expansion in the DM relative velocity, in most cases keeping only the leading term in $v^2$. 

\begin{widetext}
\begin{align} 
  \sigma v(\chi\chi\to Z^\prime\phi) &= \frac{g^4}{64\pi m_\chi^4 m^4_{Z^\prime}} \left(m_\phi^4 - 2 m_\phi^2(4 m_\chi^2+m_{Z^\prime}^2) + (4 m_\chi^2-m_{Z^\prime}^2)^2\right)^{3/2}\,, \label{eq:xsec1} \\
  \sigma v(\chi\chi\to Z^\prime Z^\prime) &= \frac{g^4}{16\pi m_\chi^2}\left(1-\frac{m_{Z^\prime}^2}{m_\chi^2}\right)^{1/2}\left(1 + \frac{4(8m_\chi^4+m_\phi^4)}{3(4m_\chi^2-m_\phi^2)^2}\frac{m_\chi^4}{m_{Z^\prime}^4}v^2 + \mathcal{O}\bigg(\frac{m_{Z^\prime}^2}{m_\chi^2}\bigg) \right) \,, \label{eq:xsec2} \\
  \sigma v(\chi\chi\to \phi\phi) &= \frac{3g^4 m_\chi^2}{8\pi m_{Z^\prime}^4}\left(1-\frac{m_\phi^2}{m_\chi^2}\right)^{1/2}\left(1+\mathcal{O}\bigg(\frac{m_\phi^2}{m_\chi^2}\bigg)\right) v^2 \,, \label{eq:xsec3} \\
  \sigma v(\chi\chi\to \bar{f}f) &= \frac{g^4 N_c Q_f^2}{12 \pi m_{\chi}^2}\left(1-\frac{m^2_f}{m_\chi^2}\right)^{1/2} \frac{(2+m_f^2/ m_{\chi}^2)}{(4-m_{Z^\prime}^2/m_\chi^2)^2}\, v^2 \,. \label{eq:xsec4}
\end{align}
\end{widetext}
In the last equation, $Q_f$ denotes the $U(1)_{(B-L)_3}$ charge of $f$, $N_c$ is the number of colours in the case of annihilation to quarks, and for annihilation to neutrinos the expression should be multiplied by $1/2$.

Notice that the only s-wave annihilation channels are $Z^\prime\phi$ and $Z^\prime Z^\prime$, and so these are expected to dominate whenever they are kinematically accessible. 
In fact, the $Z^\prime\phi$ channel always dominates, as can be clearly understood by taking the $m_\chi\gg m_{Z^\prime},\, m_\phi$ limit in Eqs.~\eqref{eq:xsec1} \& \eqref{eq:xsec2}. 
It should therefore be clear that this model cannot be described by a single mediator simplified model across much of its parameter space.

For $\chi\chi\to Z^\prime Z^\prime$, we have also included the p-wave term above, which is naively expected to be sub-leading. 
However, it is enhanced relative to the s-wave contribution by the potentially large factor $m_\chi^4/m_{Z^\prime}^4$, and hence can be important during freeze-out. 
The two remaining annihilation channels are p-wave and both can be relevant for $m_\chi<m_{Z^\prime}$, depending on the $\phi$ mass.

In practice, we do not use the above approximate expressions in our numerical analysis. 
Rather, {\tt micrOMEGAs-4.3}~\cite{1407.6129} is used to compute the relic density, as well as the cross-sections for direct and indirect detection. 
Note that {\tt micrOMEGAs} does not include the effects of $2-3$ annihilations via an off-shell $\phi$ or $Z^\prime$; however, these are expected to be important only in narrow regions of parameter space below the $2-2$ thresholds. 
Finally, recall that the gauge and Yukawa couplings are related at tree-level via Eq.~\eqref{eq:yukawa}. 
Hence, one can always choose to fix the gauge coupling in order to obtain the correct relic density, $\Omega_{DM}=0.12$.

\section{Experimental and Theoretical Constraints} \label{sec:constraints}

\subsection{Perturbativity and Unitarity}

Consistency of the model requires that perturbative unitarity is satisfied, which imposes upper limits on the various masses and/or couplings. 
Important bounds are obtained by considering $2-2$ scattering processes involving $\chi$, $\phi$, and the longitudinal polarization of the $Z^\prime$.
The strongest bounds generally come from the $J=0$ partial-wave amplitude, defined by
\begin{equation}
  a^0_{fi}(s) = \frac{1}{32\pi}\int_{-1}^{1}d(\cos\Theta)\mathcal{M}_{fi}(s,t) \,,
\end{equation}
where $\mathcal{M}_{fi}(s,t)$ is the scattering amplitude, with $s$ and $t$ the usual Mandelstam variables, and $\Theta$ the scattering angle in the center-of-mass frame. 
The perturbative unitarity bound can then be expressed as 
\begin{equation} \label{eq:unitarity}
  |\text{Re}(a_{ii}^0)|<1/2 \,.
\end{equation}

Firstly, let us consider scattering of the states $Z^\prime_LZ^\prime_L/\sqrt{2}$, $\phi\phi/\sqrt{2}$, and $Z^\prime_L\phi$, assuming\footnote{For unitarity bounds in the case of large mixing see Ref.~\cite{1606.07609}.} $\lambda_{H\Phi}\to0$.
In the high-energy limit, the $J=0$ partial-wave scattering matrix is then given by~\cite{Lee:1977yc, *Lee:1977eg}
\begin{equation}
  \lim_{\sqrt{s}\to\infty} a_{fi}^0 = -\frac{g^2m_\phi^2}{8\pi m_{Z^\prime}^2}  
  \left(\begin{array}{ccc}
  3&1&0\\
  1&3&0\\
  0&0&2
  \end{array}\right) .
\end{equation}
Imposing the unitarity bound~\eqref{eq:unitarity} on the eigenstate of $a_{fi}^0$ with the largest eigenvalue ($Z^\prime_LZ^\prime_L+\phi\phi$) leads to the bound
\begin{equation}
  m_\phi < \frac{\sqrt{\pi}m_{Z^\prime}}{g} \,.
\end{equation}

Similarly, by considering $\chi\chi\to\chi\chi$ scattering, one obtains the bounds~\cite{Chanowitz:1978mv, 1412.5660, 1510.02110, 1606.07609}
\begin{equation}
  m_\chi < \frac{\sqrt{\pi}m_{Z^\prime}}{g}\,, \qquad g<\sqrt{4\pi} \,.
\end{equation}
Using Eq.~\eqref{eq:masses}, the first bound can be rewritten simply as $y<\sqrt{8\pi}$.

The above perturbative unitarity bounds apply equally to both simplified and complete, self-consistent models. 
However, in the latter case there is an additional, stronger bound that can also be imposed if one is willing to make some mild assumptions about the UV physics. 
One reasonable assumption is that the couplings remain perturbative (i.e. do not encounter a Landau pole) up to the Planck scale. 

The one-loop RGEs are simply given by\footnote{In the RGE for $g$, we have assumed the presence of two additional scalars with $U(1)$ charges $+1$ and $+1/3$, which play a role in generating the SM Yukawa couplings~\cite{1705.03858}; however, their effect is small.}
\begin{equation}
  \beta_g = \frac{142}{27}\frac{g^3}{(4\pi)^2} \,, \qquad \beta_y = \frac{1}{(4\pi)^2}\frac{3}{2}\left( y^3-4g^2y\right) . 
\end{equation}
Neglecting for now the $g^2y$ term in $\beta_y$ (we use the full coupled RGEs in our numerical analysis), the RGEs have a simple solution, and yield the bounds
\begin{equation}
  g(\mu) < \frac{6\pi}{\sqrt{\frac{71}{3}\ln\frac{\Lambda}{\mu}}} \,, \qquad y(\mu) < \frac{4\pi}{\sqrt{3\ln\frac{\Lambda}{\mu}}} \,,
\end{equation}
where $\Lambda$ is the scale of the Landau pole. 
Taking $\Lambda=M_{Pl}$ and $\mu=1$\,TeV, gives $g<0.65$ and $y<1.2$. 
These constraints are therefore significantly stronger than those obtained from partial-wave perturbative unitarity.

\subsection{$Z$-$Z^\prime$ Mixing and Electroweak Precision}

There can be kinetic mixing between the $U(1)_Y$ and $U(1)_{(B-L)_3}$ gauge bosons, parametrised by $\epsilon$ in Eq.~\eqref{eq:lag}. 
This mixing leads to a shift in the $Z$ boson mass from its SM value, and as a result is tightly constrained by electroweak precision measurements which require $\epsilon\lesssim10^{-1}-10^{-2}$, depending on the $Z^\prime$ mass~\cite{1006.0973}. 
Given this constraint, we can neglect the kinetic mixing when computing other observables, such as the DM relic density.\footnote{We have confirmed numerically that the effects of kinetic mixing are indeed negligible whenever the electroweak precision bounds are satisfied.}

While $\epsilon$ is simply a free parameter from the point of view of the low-energy theory, it may be determined by the UV physics. 
If either $U(1)_Y$ or $U(1)_{(B-L)_3}$ is ultimately embedded into a non-abelian gauge group at some high scale (e.g. in a GUT), then the kinetic mixing necessarily vanishes above that scale.
At lower scales it is reintroduced radiatively, with the one-loop RGE given by
\begin{equation}
  (4\pi)^2\beta_\epsilon = -\frac{32}{9}g_Yg - \left(\frac{142}{27}g^2 + \frac{41}{6}g_Y^2\right)\epsilon \,.
\end{equation}
Considering only the leading (first) term above, and neglecting the running of the gauge couplings, one obtains the approximate solution
\begin{equation}
  \epsilon(\mu) = \frac{2 g_Yg}{9\pi^2} \log\left(\frac{\Lambda}{\mu}\right) ,
\end{equation}
where $\Lambda$ is the cut-off scale at which $\epsilon$ vanishes. 
It is clear that for either large $U(1)_{(B-L)_3}$ gauge couplings or high cut-off scales, $\epsilon(m_Z)$ can easily exceed the bound from electroweak precision measurements. 

Given that our model is self-consistent and in principle valid up to very high scales, in Sec.~\ref{sec:results} we show the bounds obtained by taking a GUT-scale cut-off, $\Lambda=10^{16}\,$GeV (we also use the full solution to the one-loop RGEs). 
As we shall see, this leads to a powerful constraint on the parameter space. 
However, it is important to keep in mind that, unlike the bounds discussed in the following sections, this limit is dependent on the above assumption about the UV physics. 
Modifying this assumption (i.e. taking a lower value for $\Lambda$) can significantly alleviate the bounds. 


\subsection{Higgs Measurements}

Measurements of the SM Higgs boson couplings and direct searches for Higgs invisible decays provide constraints on $h$-$\phi$ mixing. 
If new decay modes for the Higgs are kinematically allowed (i.e. $\chi\chi$, $Z^\prime Z^\prime$ or $\phi\phi$), these bounds are in fact already quite stringent. 
The additional partial decay widths are given by
\begin{align}
  \Gamma_{h\to\chi\chi} &= \frac{y^2\sin^2\theta\, m_h}{32\pi} \left( 1-\frac{4m_\chi^2}{m_h^2} \right)^{3/2} \,, \\
  \Gamma_{h\to Z^\prime Z^\prime} &= \frac{g^2\sin^2\theta\,m_h^3}{8\pi m_{Z^\prime}^2} \left( 1-\frac{4m_{Z^\prime}^2}{m_h^2} \right)^{1/2} \notag \\
  &\qquad\times\left(1-4\frac{m_{Z^\prime}^2}{m_h^2}+12\frac{m_{Z^\prime}^4}{m_h^4}\right) \,, \\
  \Gamma_{h\to\phi\phi} &= \frac{\sin^22\theta\,m_h^3}{128\pi v^2} \left(1-\frac{4m_\phi^2}{m_h^2}\right)^{1/2} \left(1+\frac{2m_\phi^2}{m_h^2}\right)^2 \notag \\
  &\qquad\times\left(\sin\theta-\frac{v}{w}\cos\theta\right)^2 \,. \label{eq:hphiphi}
\end{align}
On the other hand, the partial widths to SM final states are universally reduced by $\cos^2\theta$. 
One can then straightforwardly impose the bound from the overall Higgs signal strength into SM states as measured by ATLAS and CMS, $\mu=1.09\pm0.11$~\cite{1606.02266}. 
In the limit $m_h< 2m_\chi, 2m_{Z^\prime}, 2m_\phi$, this leads to $|\theta|<0.37$. 
More generally, the bound also depends on the sign of $\theta$ via Eq.~\eqref{eq:hphiphi}.

There are also direct searches for invisible decays of the Higgs~\cite{1509.00672, 1610.09218}, which lead to the bound
\begin{equation}
  \text{BR}(h\to\chi\chi)<\frac{0.24}{\cos^2\theta} \,.
\end{equation}
However, this is always weaker than the above limit from the signal strength into SM final states.

\subsection{Flavour Observables}

The new $Z^\prime$ gauge boson develops flavour off-diagonal couplings after rotation of the SM fermions to the mass basis. 
Hence, it will generically mediate FCNC at tree-level; these are strongly constrained by flavour observables. 
Generally, the strongest bounds come from $K-\bar{K}$ mixing; however, these are ameliorated within our framework by construction, due to the universal $U(1)$ charges of the first and second generation quarks. 
The effects on other flavour observables strongly depend on the detailed structure of the rotation matrices, and in particular the new physical mixing angles involving the third generation. 
A complete discussion is beyond the scope of this work and we refer the reader to Refs.~\cite{1705.03858, 1705.01822, 1705.00915} for further details. 

Instead let us focus on the most minimal case, where all new mixing angles vanish. In this case, off-diagonal couplings in the quark sector can be confined to the LH up-type quarks; this leads to constraints from $D^0-\bar{D^0}$ mixing. However, a stronger bound is obtained by considering only the flavour-diagonal couplings, which lead to effects in $\Upsilon$ decays. 
In particular, these modify the lepton universality ratio
\begin{equation}
  R_{\tau\mu} = \frac{BR(\Upsilon(1S)\to\tau^+\tau^-)}{BR(\Upsilon(1S)\to\mu^+\mu^-)} \,,
\end{equation}  
which has been precisely measured by BaBar~\cite{1002.4358}: $R_{\tau\mu}=1.005\pm0.026$. 
In our model, and neglecting the small contribution from the SM $Z$ boson, we obtain
\begin{equation}
  R_{\tau\mu} = \left( 1 + \frac{g^2}{m_{Z^\prime}^2 - m_\Upsilon^2} \frac{m_\Upsilon^2}{4\pi\alpha} \right)^2 , 
\end{equation}
up to small corrections from $m_\tau$ which are included in our numerical analysis.

\subsection{DM Direct Detection}

DM-nucleon scattering mediated by the $Z^\prime$ is highly suppressed in this model. 
Firstly, the Majorana nature of the DM leads to spin-independent scattering that is velocity-suppressed.
There is then further suppression due to the fact that the $Z^\prime$ does not couple to the light quarks. 
Although such couplings will be introduced during the RGE running down to nuclear scales, the resulting direct detection bounds are essentially negligible~\cite{1605.04917}. 
The $Z^\prime$ may also develop couplings to the light quarks after rotation to the mass basis. 
If the resulting couplings are chiral, there is then spin-dependent scattering that is no longer momentum-suppressed. 
However, the relevant mixing angles are in most cases required to be small due to flavour constraints, and the resulting direct detection bounds due to the $Z^\prime$ mediator are still expected to be very weak. 
 
The situation changes once we include non-zero $h$-$\phi$ mixing. 
The scalars then mediate spin-independent DM-nucleon scattering that is already strongly constrained by existing direct detection experiments. 
In our analysis, we impose the latest bounds from the Xenon~1T experiment~\cite{1705.06655}. 
As we shall see, these bounds give constraints even for very small mixing angles ($\theta\sim10^{-3}$). 
Furthermore, an important point to note is that such mixing is at the very least generated radiatively at the two-loop level, i.e. $\lambda_{H \Phi}\sim y_t^2g^4/(4\pi)^4$; hence the mixing angle cannot simply be taken to be arbitrarily small.

\subsection{DM Indirect Detection}

In regions of parameter space where the dominant annihilation channel is either $\chi\chi\to Z^\prime\phi$ or $\chi\chi\to Z^\prime Z^\prime$, the annihilation is s-wave and there can be potentially observable indirect detection signals. 
However, as discussed in Sec.~\ref{sec:relic_density}, the p-wave cross-section is enhanced for $\chi\chi\to Z^\prime Z^\prime$ and contributes in setting the relic density in the early universe. 
As a result, the annihilation cross-section today can be significantly lower than the thermal relic cross-section. 
When considering the existing bounds from indirect detection experiments, we can therefore restrict ourselves to regions of parameter space where $\chi\chi\to Z^\prime\phi$ is the dominant annihilation channel. 

The strongest indirect detection limits on this model come from the Fermi-LAT gamma ray observations of dark matter dominated, dwarf-spheroidal galaxies~\cite{1611.03184}. 
We generate gamma ray spectra from $\chi\chi\to Z^\prime\phi$ annihilation for a range of DM, $Z^\prime$, and $\phi$ masses using {\tt Pythia-6.4}~\cite{hep-ph/0603175}. 
The Fermi-LAT binned likelihood functions are then used to compute 95\%~C.L. upper limits on the DM annihilation cross-section for each dwarf galaxy. 
The statistical uncertainties on the J-factors are included as nuisance parameters, following the statistical method adopted in Ref.~\cite{1503.02641}. 
For our overall limit, we simply take the strongest limit obtained from the "nominal" set of dwarf galaxies in Ref.~\cite{1611.03184}; in practice this is always from either Ursa Major II or Ursa Minor.

\subsection{LHC Searches}

There are in general two ways to probe dark matter models in collider searches. 
Firstly, one can attempt to directly probe the dark sector via searches for large missing energy, in particular mono-jet searches. 
Alternatively, one can look for the mediators via their decays into SM final states. 
Within this model, the $Z^\prime$ coupling to leptons means that the second approach will provide the most sensitive bounds.  

An important aspect of this model for LHC searches is the absence of $Z^\prime$ couplings to the first and second generation quarks. 
This leads to a significantly suppressed production cross-section compared to generic $Z^\prime$ models. 
Even if such couplings are introduced after rotation to the mass basis, the strong constraints on the mixing angles from flavour processes ensure that $\bar{b}b\to Z^\prime$ is expected to remain the dominant production channel. 
In determining the LHC bounds, the production cross-section is calculated in the 5-flavour scheme at NLO using {\tt MadGraph-2.5.4}~\cite{1405.0301}.

As a result of the suppressed production cross-section, mono-jet searches~\cite{1703.01651, ATLAS-CONF-2017-060} are not currently sensitive to this model. 
On the other hand, searches for the $Z^\prime$ in leptonic final states can provide good sensitivity. 
The precise bounds again depend on the $Z^\prime$ couplings in the mass basis; we shall assume the minimal case where there is no rotation of the charged leptons. 
The LHC bound then comes from $Z^\prime\to\bar{\tau}\tau$ searches. 
We impose the limits from the recent ATLAS search for spin-1 resonances with $36.1\,\text{fb}^{-1}$ integrated luminosity at $\sqrt{s}=13\,$TeV~\cite{1709.07242}. 

Finally, one can also search for the scalar $\phi$. 
It has a potentially rich phenomenology, and can decay to $\chi\chi$, $Z^\prime Z^\prime$, $hh$, and other SM final states via mixing. 
Each of these decays can dominate in certain regions of parameter space. 
However, $\phi$ production at the LHC is always suppressed by the mixing angle $\theta$. 
After taking into account the other bounds on the mixing angle discussed above, the production cross-section is expected to be small; hence we shall not consider the detailed bounds from these searches.

\section{Limits on the Parameter Space} \label{sec:results}

The constraints on the model parameter space from the bounds discussed in Sec.~\ref{sec:constraints} are combined in Figs.~\ref{fig:m13} \& \ref{fig:m3}. 
Results are shown in the $m_\chi$-$m_{Z^\prime}$ plane, where we have fixed the gauge coupling by requiring that the observed DM relic density is satisfied. 
We shall consider two values for $m_\phi$, which exemplify the two qualitatively different regions of parameter space.\footnote{The resonance case, $m_\phi\approx2m_\chi$, is also qualitatively distinct; however, we will not consider this special case in detail.} 
Similarly, we consider both large ($\theta=0.1$) and small ($\theta=10^{-3}$) mixing.

\subsection{$m_\phi=m_\chi/3$}

\begin{figure*}[t]
  \centering
  \includegraphics[width=0.45\textwidth]{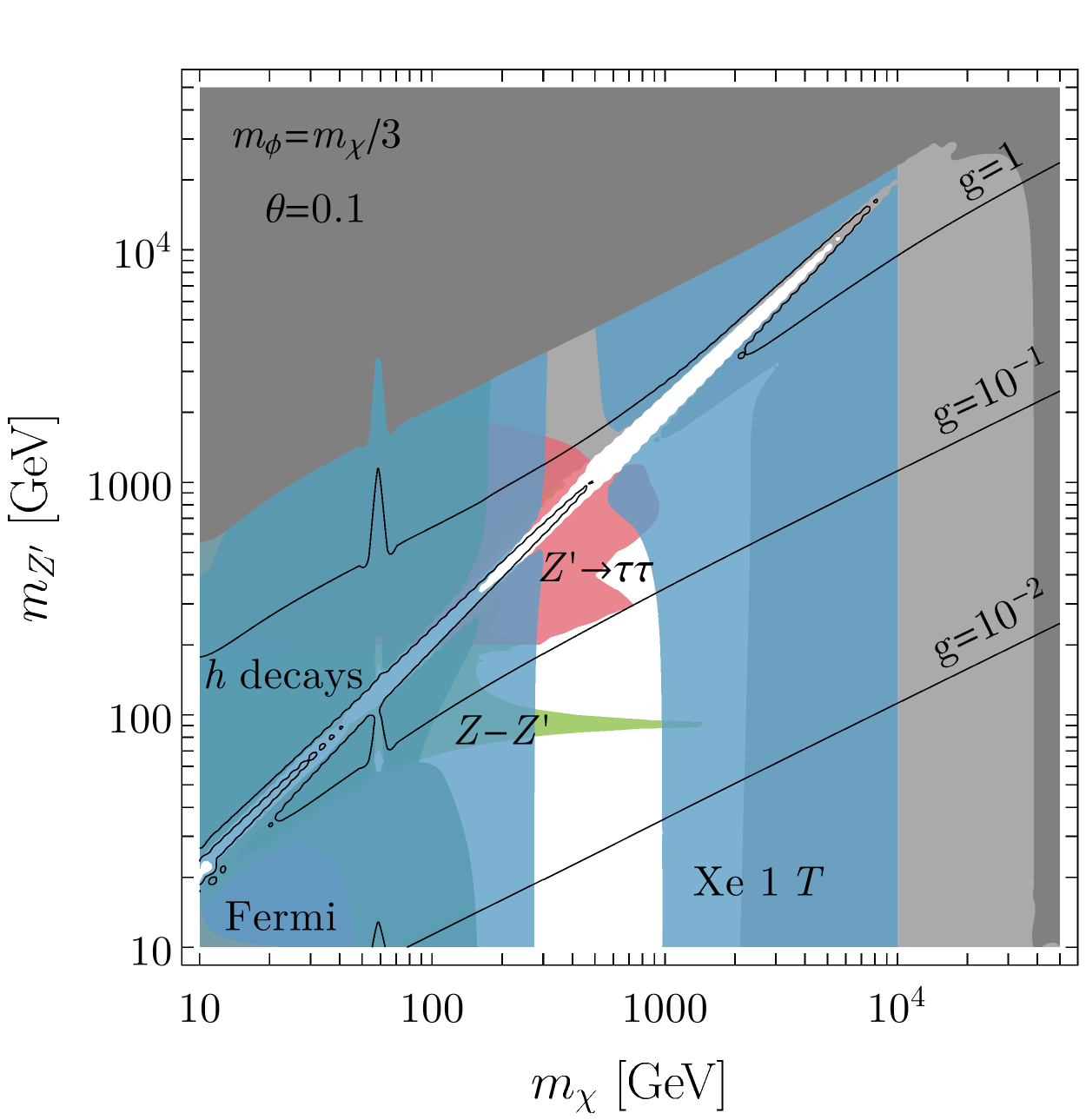}
  \includegraphics[width=0.45\textwidth]{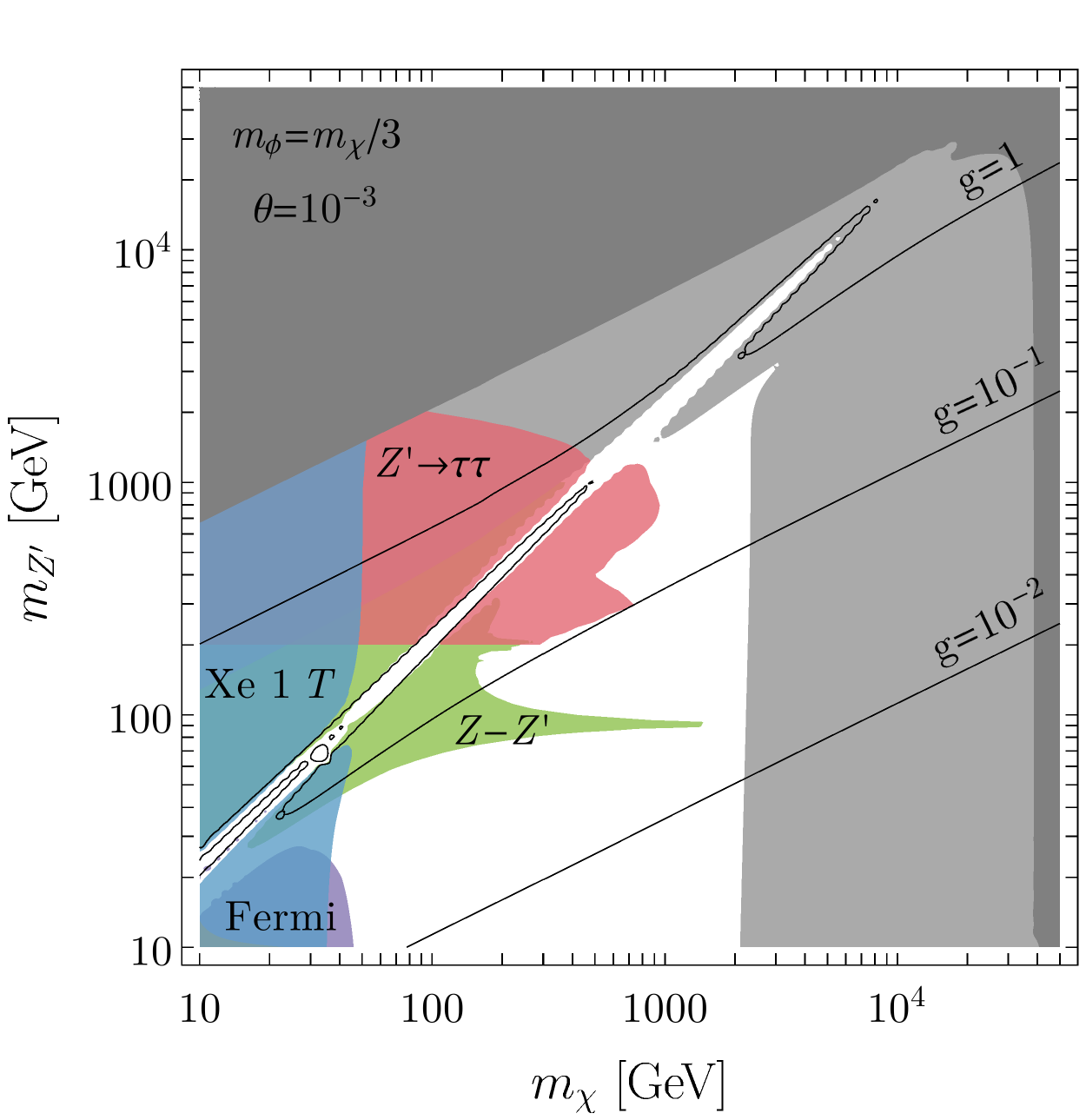} 
  \caption{Constraints on the model shown in the $m_\chi$-$m_{Z^\prime}$ plane, in the case of large mixing (left) and small mixing (right). All regions satisfy the correct relic density, with the contours showing the required gauge coupling. The dark grey regions violate perturbative unitarity, while in the light grey regions the gauge or Yukawa couplings encounter a Landau pole below $M_{Pl}$. The coloured regions are excluded by various experimental measurements at 95\% C.L. (90\% C.L. for Xenon-1T). We have fixed $m_\phi=m_\chi/3$.}
  \label{fig:m13}
\end{figure*}

In Fig.~\ref{fig:m13} we fix $m_\phi=m_\chi/3$ such that the annihilation channel $\chi\chi\to Z^\prime\phi$ is kinematically accessible, provided $m_\chi\gtrsim3m_{Z^\prime}/5$.  
Firstly, notice the clearly visible region where the annihilation cross-section is resonantly enhanced, $m_\chi\approx m_{Z^\prime}/2$. 
To the right of this region $\chi\chi\to Z^\prime\phi$ is always the dominant annihilation channel, while for smaller DM masses both $\chi\chi\to\bar{f}{f}$ and $\chi\chi\to\phi\phi$ are relevant. 
In the case of large mixing (left panel), Higgs-mediated annihilation can also be important near resonance, $m_\chi\approx m_h/2$.

The dark grey region in Fig.~\ref{fig:m13} shows where partial-wave perturbative unitarity is violated; this bound is also obtained in simplified models.  
On the other hand, the light grey region shows the much stronger constraint that can be obtained in the complete model, if one requires that all couplings remain perturbative up to the Planck scale. 
In particular, the maximum dark matter mass is significantly reduced from $m_\chi\approx40$\,TeV to $m_\chi\approx2$\,TeV (excepting the resonance region). 
Notice that for a given value of $m_\chi$ above the resonance region, the relic density requirement fixes $w=m_{Z^\prime}/(2g)$ via Eq.~\eqref{eq:xsec1}; hence, the perturbativity bound is simply a vertical line where the required Yukawa coupling becomes too large ($y(m_{Z^\prime})\gtrsim1.2$). 
There is also an upper bound on the $Z^\prime$ mass coming from the requirement that the gauge coupling remains perturbative ($g(m_{Z^\prime})\lesssim0.65$).

Let us now turn to the various experimental bounds, denoted by the coloured regions in Fig.~\ref{fig:m13}. 
Firstly, note that large $h$-$\phi$ mixing ($\theta=0.1$) is already highly constrained by the results from the Xenon-1T experiment.  
The region that survives is where $m_\phi\approx m_h$; in this case there is a cancellation between the two contributions to the DM-nucleon scattering cross-section. 
The resonance region ($m_\chi\approx m_{Z^\prime}/2$) also partially survives due to the smaller gauge coupling. 
For DM masses below $\sim100$\,GeV, there is also a strong bound from Higgs measurements; however, this is always weaker than the bound from direct detection. 

Even in the case of very small mixing ($\theta=10^{-3}$), we see that there are still non-negligible constraints from direct detection. 
However, there are now also large regions of viable parameter space. 
For $m_{Z^\prime}\sim m_Z$ there is a relatively strong bound from electroweak precision measurements due to the kinetic mixing; however, recall that this bound can be significantly weakened by modifying our assumptions regarding the UV physics. 
LHC searches also play an important role, and provide the best way to probe this model for large $Z^\prime$ masses. 
In fact, a naive rescaling of the current bounds to $3000\,\text{fb}^{-1}$ suggests that, for $m_{Z^\prime}\gtrsim200$\,GeV, the HL-LHC will be able to probe the entire region up to the perturbativity bound ($m_\chi\lesssim2$\,TeV).
For small $Z^\prime$ masses, there are constraints from indirect detection. 
While the current bounds from Fermi-LAT are relatively weak, future indirect detection experiments such as CTA will be essential for probing this region of parameter space, and will be sensitive to heavier DM masses~\cite{1508.06128}.

\subsection{$m_\phi=3m_\chi$}

\begin{figure*}[t]
  \centering
  \includegraphics[width=0.45\textwidth]{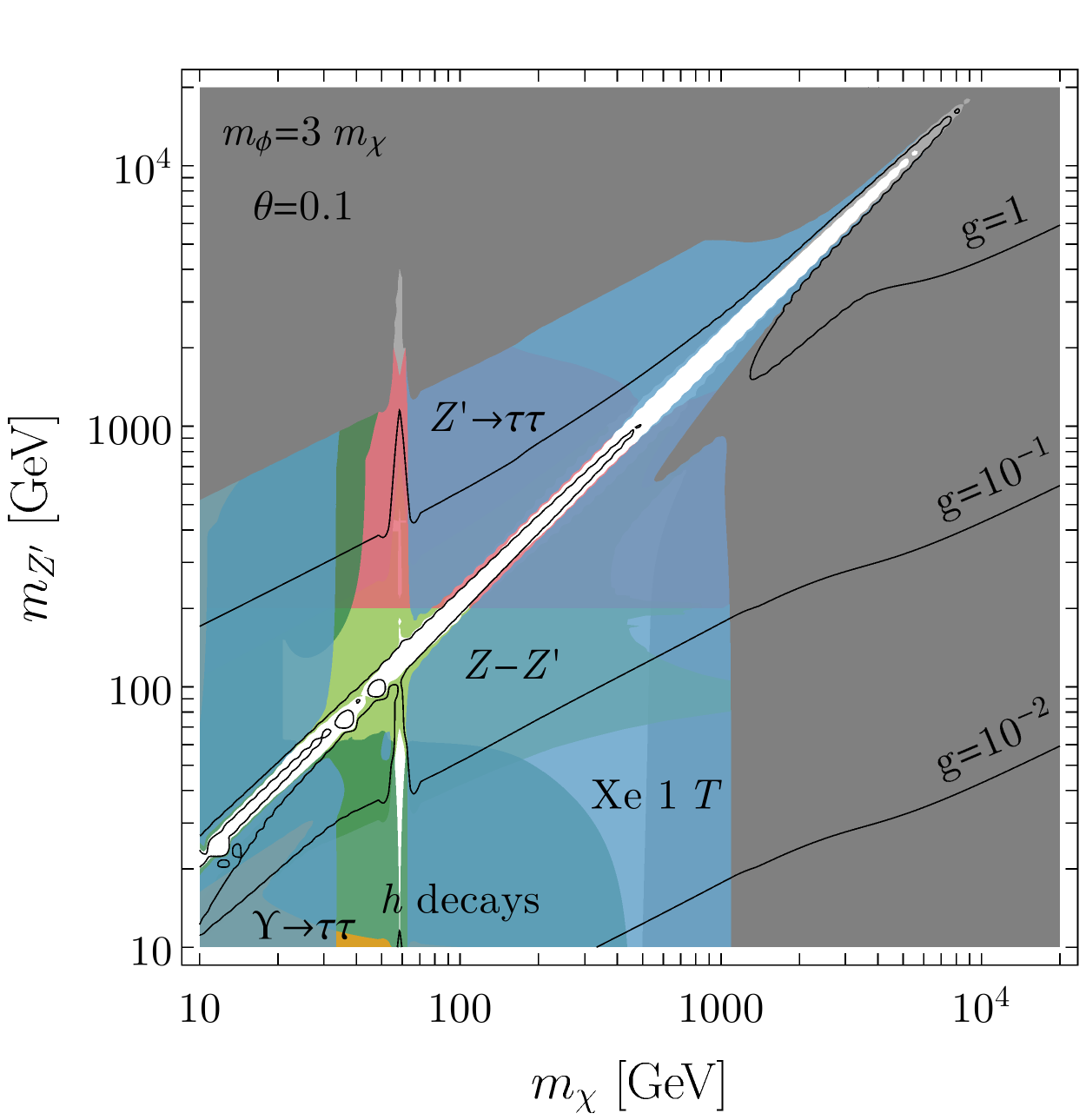}
  \includegraphics[width=0.45\textwidth]{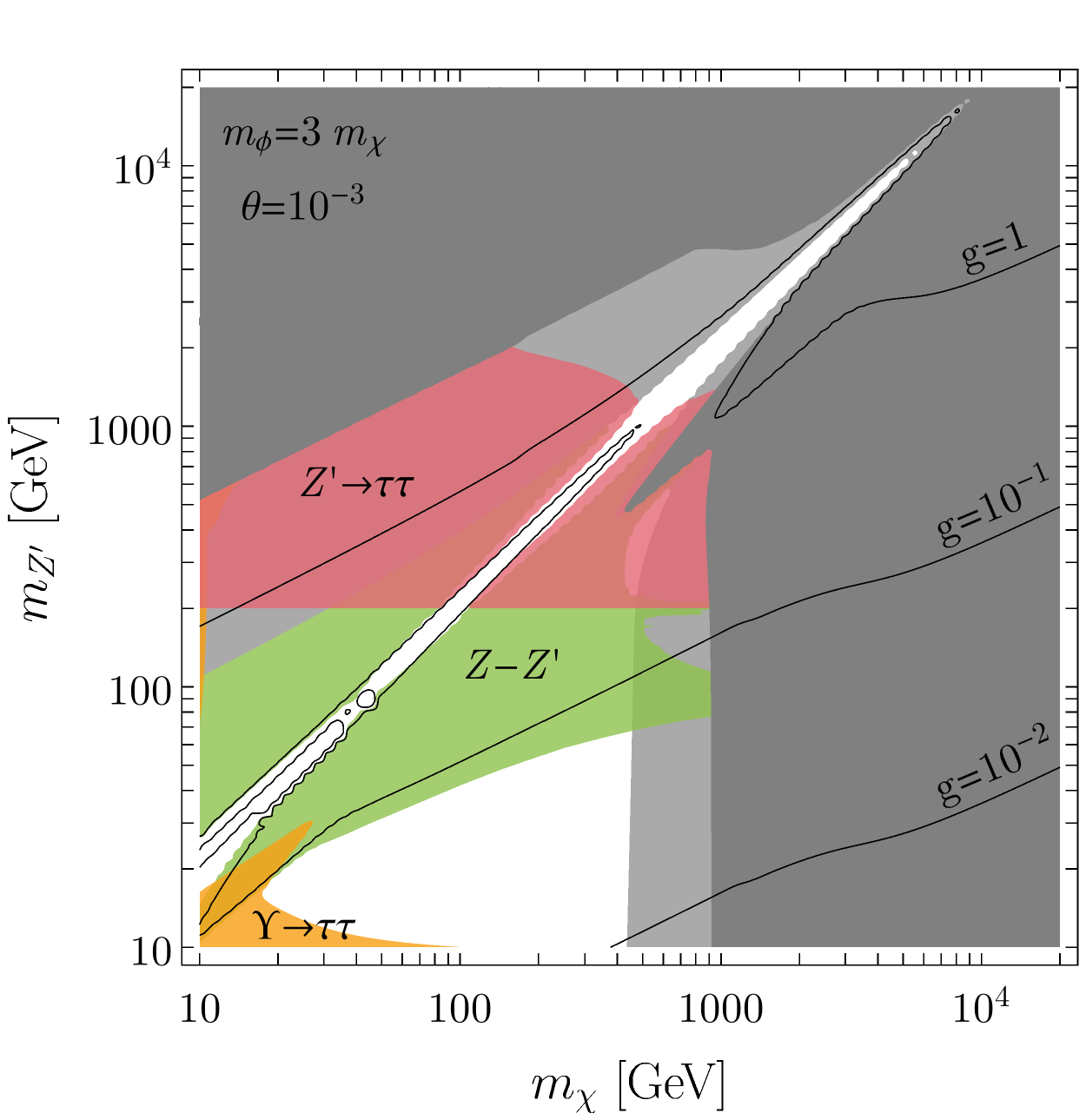}
  \caption{As in Fig.~\ref{fig:m13}, but with $m_\phi=3m_\chi$}
  \label{fig:m3}
\end{figure*}

Next, we consider the case where $m_\phi=3m_\chi$ and the $\chi\chi\to Z^\prime\phi$ annihilation channel is always kinematically forbidden; this is shown in Fig.~\ref{fig:m3}.
A key difference compared to the previous case is that freeze-out proceeds via $\chi\chi\to Z^\prime Z^\prime$ annihilation for DM masses above the resonance region. 
Consequently, a larger gauge coupling is required in order to achieve the thermal relic cross-section.
As a result, this region of parameter space becomes much more strongly constrained. 
Note that for large mixing, $\chi\chi\to h Z^\prime$ is now also an important annihilation channel for heavy DM masses; however, the suppression by the mixing angle means that increased gauge couplings are still needed.

For large dark matter masses, the p-wave term in Eq.~\eqref{eq:xsec2} dominates and the relic density requirement again fixes $w=m_{Z^\prime}/(2g)$. 
Since the required value of $w$ is smaller than in the $m_\phi=m_\chi/3$ case, the upper bound on the dark matter mass from perturbativity of the Yukawa coupling becomes significantly stronger. 
Outside of the resonance region, partial-wave unitarity now requires $m_\chi\lesssim1$\,TeV and perturbativity up to the Planck scale $m_\chi\lesssim0.5$\,TeV. 
On the other hand, for DM masses below $m_{Z^\prime}/2$ the situation is largely unchanged from before (although the $\chi\chi\to\phi\phi$ channel is now forbidden), and as such the upper bound on the $Z^\prime$ mass is similar. 

From the left panel of Fig.~\ref{fig:m3}, we see that large mixing is almost completely excluded by the various experimental measurements, with the exception of the resonance regions $m_\chi\approx m_{Z^\prime}/2$ and $m_\chi\approx m_h/2$. 
The main reason for this is that the condition $m_\phi\approx m_h$, where the bounds from direct detection can be evaded, is now satisfied for smaller DM masses where there are already constraints from other measurements. 

Moving to the right panel of Fig.~\ref{fig:m3}, we see that even the small mixing case is quite strongly constrained. 
LHC searches benefit from the larger gauge coupling and cover most of the allowed parameter space for $m_{Z^\prime}>200$\,GeV. 
Similarly, the kinetic mixing parameter is sensitive to the gauge coupling via the RGE running, and electroweak precision measurements exclude much of the intermediate $Z^\prime$ mass region (once again this bound is UV sensitive). 
For a light $Z^\prime$, there is now also a constraint from $\Upsilon$ decays. 

Interestingly, the bound from direct detection is actually alleviated due to the heavier $\phi$ mass, which reduces the DM-nucleon scattering cross-section. 
Finally, there are no bounds from indirect detection since the s-wave annihilation cross-section today can be significantly smaller than the thermal relic cross-section, as discussed in Sec.~\ref{sec:relic_density}. 
The low $Z^\prime$ mass region is therefore challenging to probe via dark matter searches in this case; however, such a $Z^\prime$ could be probed at a future $e+$\,$e-$ collider~\cite{1704.00730}.

\section{Conclusion} \label{sec:conclusion}

We have considered the possibility that the lightest RH neutrino could provide the observed dark matter in the universe. 
In particular, we assumed this RH neutrino is charged under a new $U(1)$ gauge symmetry that is spontaneously broken near the TeV scale. 
It can then obtain a Majorana mass upon $U(1)$ breaking, and be produced via thermal freeze-out, making it a standard WIMP dark matter candidate.

Interestingly, we have shown that if one requires two additional heavy RH neutrinos for leptogenesis, and demands suppression of FCNC in the quark sector, then there remains only a single-parameter family of possible vectorial $U(1)$ symmetries. 
We then focused in detail on the special case of a flavoured $B-L$ symmetry, which is the least constrained model within this framework. 

In the case of large mixing ($\theta=0.1$) between the $U(1)_{(B-L)_3}$ breaking scalar and the SM Higgs, this model is already highly constrained by DM direct detection experiments. 
For smaller mixing angles, these bounds are alleviated and there remains significant parameter space to be explored. 
The region $m_{Z^\prime}>200$\,GeV can be probed via $Z^\prime$ searches at the LHC. 
For lighter $Z^\prime$ masses, certain regions of parameter space can be tested with future indirect detection experiments; however, the situation is more challenging.

Finally, if one requires consistency of the model up to the Planck scale, there are strong upper limits on the $Z^\prime$ and, in particular, the DM masses. 
Outside of resonance regions, we obtain the bound $m_\chi\lesssim2$\,TeV. 
This limit is significantly stronger than that obtained by imposing perturbative unitarity in simplified models.

\vspace{5ex}

\noindent {\bf{Acknowledgements}}
This work is supported by Grants-in-Aid for Scientific Research from the Ministry of Education, Culture, Sports, Science, and Technology (MEXT), Japan, No. 26104009 (T.T.Y.), No. 16H02176 (T.T.Y.) and No. 17H02878 (T.T.Y.), and by the World Premier International Research Center Initiative (WPI), MEXT, Japan (P.C., C.H. and T.T.Y.). 

\vspace{5ex}

\appendix

\section{$h$-$\phi$ Mixing} \label{app:mixing}

Let us define the mass eigenstates $h^\prime$, $\phi^\prime$. The gauge eigenstates can then be expressed as
\begin{align}
  h &= h^\prime\cos\theta + \phi^\prime\sin\theta \,, \notag \\
  \phi &= \phi^\prime\cos\theta - h^\prime\sin\theta \,,
\end{align}
where the mixing angle $\theta\in[-\pi/4,\,\pi/4]$ is given by
\begin{equation}
  \tan 2\theta = \frac{\lambda_{H \Phi} v w}{\lambda_\Phi w^2 - \lambda_H v^2} \,.
\end{equation}
It is straightforward to determine the physical masses
\begin{align}
  m_{h,\phi}^2 &= \lambda_H v^2 + \lambda_\Phi w^2 \notag \\
              &\qquad\pm \left( \lambda_H v^2 - \lambda_\Phi w^2 \right)\sqrt{1+\tan^22\theta} \,,
\end{align}
where $m_h$ ($m_\phi$) is defined to be the mass of the Higgs-like state $h^\prime$ ($\phi$-like state $\phi^\prime$). 
Finally, the quartic couplings can be conveniently expressed in terms of the mixing angle and masses as
\begin{align} \label{eq:quartics}
  \lambda_H &= \frac{1}{4v^2}\left( m_\phi^2 + m_h^2 - (m_\phi^2 - m_h^2)\cos2\theta \right) , \notag \\
  \lambda_\Phi &= \frac{1}{4 w^2}\left( m_\phi^2 + m_h^2 + (m_\phi^2 - m_h^2)\cos2\theta \right) , \notag \\
  \lambda_{H \Phi} &= \frac{1}{2 w v}(m_\phi^2 - m_h^2)\sin2\theta \,.
\end{align}

\bibliographystyle{apsrev4-1-mod.bst}
\bibliography{vR-DM}

\begin{thebibliography}{45}%
\makeatletter
\providecommand \@ifxundefined [1]{%
 \@ifx{#1\undefined}
}%
\providecommand \@ifnum [1]{%
 \ifnum #1\expandafter \@firstoftwo
 \else \expandafter \@secondoftwo
 \fi
}%
\providecommand \@ifx [1]{%
 \ifx #1\expandafter \@firstoftwo
 \else \expandafter \@secondoftwo
 \fi
}%
\providecommand \natexlab [1]{#1}%
\providecommand \enquote  [1]{``#1''}%
\providecommand \bibnamefont  [1]{#1}%
\providecommand \bibfnamefont [1]{#1}%
\providecommand \citenamefont [1]{#1}%
\providecommand \href@noop [0]{\@secondoftwo}%
\providecommand \href [0]{\begingroup \@sanitize@url \@href}%
\providecommand \@href[1]{\@@startlink{#1}\@@href}%
\providecommand \@@href[1]{\endgroup#1\@@endlink}%
\providecommand \@sanitize@url [0]{\catcode `\\12\catcode `\$12\catcode
  `\&12\catcode `\#12\catcode `\^12\catcode `\_12\catcode `\%12\relax}%
\providecommand \@@startlink[1]{}%
\providecommand \@@endlink[0]{}%
\providecommand \url  [0]{\begingroup\@sanitize@url \@url }%
\providecommand \@url [1]{\endgroup\@href {#1}{\urlprefix }}%
\providecommand \urlprefix  [0]{URL }%
\providecommand \Eprint [0]{\href }%
\providecommand \doibase [0]{http://dx.doi.org/}%
\providecommand \selectlanguage [0]{\@gobble}%
\providecommand \bibinfo  [0]{\@secondoftwo}%
\providecommand \bibfield  [0]{\@secondoftwo}%
\providecommand \translation [1]{[#1]}%
\providecommand \BibitemOpen [0]{}%
\providecommand \bibitemStop [0]{}%
\providecommand \bibitemNoStop [0]{.\EOS\space}%
\providecommand \EOS [0]{\spacefactor3000\relax}%
\providecommand \BibitemShut  [1]{\csname bibitem#1\endcsname}%
\let\auto@bib@innerbib\@empty
\bibitem [{\citenamefont {Minkowski}(1977)}]{Minkowski:1977sc}%
  \BibitemOpen
  \bibfield  {author} {\bibinfo {author} {\bibfnamefont {P.}~\bibnamefont
  {Minkowski}},\ }\href {\doibase 10.1016/0370-2693(77)90435-X} {\bibfield
  {journal} {\bibinfo  {journal} {Phys. Lett.}\ }\textbf {\bibinfo {volume}
  {67B}},\ \bibinfo {pages} {421} (\bibinfo {year} {1977})}\BibitemShut
  {NoStop}%
\bibitem [{\citenamefont {Yanagida}(1979)}]{Yanagida:1979as}%
  \BibitemOpen
  \bibfield  {author} {\bibinfo {author} {\bibfnamefont {T.}~\bibnamefont
  {Yanagida}},\ }\bibfield  {booktitle} {\emph {\bibinfo {booktitle}
  {{Proceedings: Workshop on the Unified Theories and the Baryon Number in the
  Universe: Tsukuba, Japan, February 13-14, 1979}}},\ }\href@noop {} {\bibfield
   {journal} {\bibinfo  {journal} {Conf. Proc.}\ }\textbf {\bibinfo {volume}
  {C7902131}},\ \bibinfo {pages} {95} (\bibinfo {year} {1979})}\BibitemShut
  {NoStop}%
\bibitem [{\citenamefont {Glashow}(1980)}]{Glashow:1979nm}%
  \BibitemOpen
  \bibfield  {author} {\bibinfo {author} {\bibfnamefont {S.~L.}\ \bibnamefont
  {Glashow}},\ }\bibfield  {booktitle} {\emph {\bibinfo {booktitle} {{Cargese
  Summer Institute: Quarks and Leptons Cargese, France, July 9-29, 1979}}},\
  }\href {\doibase 10.1007/978-1-4684-7197-7_15} {\bibfield  {journal}
  {\bibinfo  {journal} {NATO Sci. Ser. B}\ }\textbf {\bibinfo {volume} {61}},\
  \bibinfo {pages} {687} (\bibinfo {year} {1980})}\BibitemShut {NoStop}%
\bibitem [{\citenamefont {Gell-Mann}\ \emph {et~al.}(1979)\citenamefont
  {Gell-Mann}, \citenamefont {Ramond},\ and\ \citenamefont
  {Slansky}}]{GellMann:1980vs}%
  \BibitemOpen
  \bibfield  {author} {\bibinfo {author} {\bibfnamefont {M.}~\bibnamefont
  {Gell-Mann}}, \bibinfo {author} {\bibfnamefont {P.}~\bibnamefont {Ramond}}, \
  and\ \bibinfo {author} {\bibfnamefont {R.}~\bibnamefont {Slansky}},\
  }\bibfield  {booktitle} {\emph {\bibinfo {booktitle} {{Supergravity Workshop
  Stony Brook, New York, September 27-28, 1979}}},\ }\href@noop {} {\bibfield
  {journal} {\bibinfo  {journal} {Conf. Proc.}\ }\textbf {\bibinfo {volume}
  {C790927}},\ \bibinfo {pages} {315} (\bibinfo {year} {1979})},\ \Eprint
  {http://arxiv.org/abs/1306.4669} {arXiv:1306.4669 [hep-th]} \BibitemShut
  {NoStop}%
\bibitem [{\citenamefont {Fukugita}\ and\ \citenamefont
  {Yanagida}(1986)}]{Fukugita:1986hr}%
  \BibitemOpen
  \bibfield  {author} {\bibinfo {author} {\bibfnamefont {M.}~\bibnamefont
  {Fukugita}}\ and\ \bibinfo {author} {\bibfnamefont {T.}~\bibnamefont
  {Yanagida}},\ }\href {\doibase 10.1016/0370-2693(86)91126-3} {\bibfield
  {journal} {\bibinfo  {journal} {Phys. Lett.}\ }\textbf {\bibinfo {volume}
  {B174}},\ \bibinfo {pages} {45} (\bibinfo {year} {1986})}\BibitemShut
  {NoStop}%
\bibitem [{\citenamefont {Asaka}\ and\ \citenamefont
  {Shaposhnikov}(2005)}]{hep-ph/0505013}%
  \BibitemOpen
  \bibfield  {author} {\bibinfo {author} {\bibfnamefont {T.}~\bibnamefont
  {Asaka}}\ and\ \bibinfo {author} {\bibfnamefont {M.}~\bibnamefont
  {Shaposhnikov}},\ }\href {\doibase 10.1016/j.physletb.2005.06.020} {\bibfield
   {journal} {\bibinfo  {journal} {Phys. Lett.}\ }\textbf {\bibinfo {volume}
  {B620}},\ \bibinfo {pages} {17} (\bibinfo {year} {2005})},\ \Eprint
  {http://arxiv.org/abs/hep-ph/0505013} {arXiv:hep-ph/0505013 [hep-ph]}
  \BibitemShut {NoStop}%
\bibitem [{\citenamefont {Dodelson}\ and\ \citenamefont
  {Widrow}(1994)}]{hep-ph/9303287}%
  \BibitemOpen
  \bibfield  {author} {\bibinfo {author} {\bibfnamefont {S.}~\bibnamefont
  {Dodelson}}\ and\ \bibinfo {author} {\bibfnamefont {L.~M.}\ \bibnamefont
  {Widrow}},\ }\href {\doibase 10.1103/PhysRevLett.72.17} {\bibfield  {journal}
  {\bibinfo  {journal} {Phys. Rev. Lett.}\ }\textbf {\bibinfo {volume} {72}},\
  \bibinfo {pages} {17} (\bibinfo {year} {1994})},\ \Eprint
  {http://arxiv.org/abs/hep-ph/9303287} {arXiv:hep-ph/9303287 [hep-ph]}
  \BibitemShut {NoStop}%
\bibitem [{\citenamefont {Kusenko}\ \emph {et~al.}(2010)\citenamefont
  {Kusenko}, \citenamefont {Takahashi},\ and\ \citenamefont
  {Yanagida}}]{1006.1731}%
  \BibitemOpen
  \bibfield  {author} {\bibinfo {author} {\bibfnamefont {A.}~\bibnamefont
  {Kusenko}}, \bibinfo {author} {\bibfnamefont {F.}~\bibnamefont {Takahashi}},
  \ and\ \bibinfo {author} {\bibfnamefont {T.~T.}\ \bibnamefont {Yanagida}},\
  }\href {\doibase 10.1016/j.physletb.2010.08.031} {\bibfield  {journal}
  {\bibinfo  {journal} {Phys. Lett.}\ }\textbf {\bibinfo {volume} {B693}},\
  \bibinfo {pages} {144} (\bibinfo {year} {2010})},\ \Eprint
  {http://arxiv.org/abs/1006.1731} {arXiv:1006.1731 [hep-ph]} \BibitemShut
  {NoStop}%
\bibitem [{\citenamefont {Frampton}\ \emph {et~al.}(2002)\citenamefont
  {Frampton}, \citenamefont {Glashow},\ and\ \citenamefont
  {Yanagida}}]{hep-ph/0208157}%
  \BibitemOpen
  \bibfield  {author} {\bibinfo {author} {\bibfnamefont {P.~H.}\ \bibnamefont
  {Frampton}}, \bibinfo {author} {\bibfnamefont {S.~L.}\ \bibnamefont
  {Glashow}}, \ and\ \bibinfo {author} {\bibfnamefont {T.}~\bibnamefont
  {Yanagida}},\ }\href {\doibase 10.1016/S0370-2693(02)02853-8} {\bibfield
  {journal} {\bibinfo  {journal} {Phys. Lett.}\ }\textbf {\bibinfo {volume}
  {B548}},\ \bibinfo {pages} {119} (\bibinfo {year} {2002})},\ \Eprint
  {http://arxiv.org/abs/hep-ph/0208157} {arXiv:hep-ph/0208157 [hep-ph]}
  \BibitemShut {NoStop}%
\bibitem [{\citenamefont {Ibe}\ \emph {et~al.}(2016)\citenamefont {Ibe},
  \citenamefont {Kusenko},\ and\ \citenamefont {Yanagida}}]{1602.03003}%
  \BibitemOpen
  \bibfield  {author} {\bibinfo {author} {\bibfnamefont {M.}~\bibnamefont
  {Ibe}}, \bibinfo {author} {\bibfnamefont {A.}~\bibnamefont {Kusenko}}, \ and\
  \bibinfo {author} {\bibfnamefont {T.~T.}\ \bibnamefont {Yanagida}},\ }\href
  {\doibase 10.1016/j.physletb.2016.05.025} {\bibfield  {journal} {\bibinfo
  {journal} {Phys. Lett.}\ }\textbf {\bibinfo {volume} {B758}},\ \bibinfo
  {pages} {365} (\bibinfo {year} {2016})},\ \Eprint
  {http://arxiv.org/abs/1602.03003} {arXiv:1602.03003 [hep-ph]} \BibitemShut
  {NoStop}%
\bibitem [{\citenamefont {Alonso}\ \emph
  {et~al.}(2017{\natexlab{a}})\citenamefont {Alonso}, \citenamefont {Cox},
  \citenamefont {Han},\ and\ \citenamefont {Yanagida}}]{1704.08158}%
  \BibitemOpen
  \bibfield  {author} {\bibinfo {author} {\bibfnamefont {R.}~\bibnamefont
  {Alonso}}, \bibinfo {author} {\bibfnamefont {P.}~\bibnamefont {Cox}},
  \bibinfo {author} {\bibfnamefont {C.}~\bibnamefont {Han}}, \ and\ \bibinfo
  {author} {\bibfnamefont {T.~T.}\ \bibnamefont {Yanagida}},\ }\href {\doibase
  10.1103/PhysRevD.96.071701} {\bibfield  {journal} {\bibinfo  {journal} {Phys.
  Rev.}\ }\textbf {\bibinfo {volume} {D96}},\ \bibinfo {pages} {071701}
  (\bibinfo {year} {2017}{\natexlab{a}})},\ \Eprint
  {http://arxiv.org/abs/1704.08158} {arXiv:1704.08158 [hep-ph]} \BibitemShut
  {NoStop}%
\bibitem [{\citenamefont {Bonilla}\ \emph {et~al.}(2017)\citenamefont
  {Bonilla}, \citenamefont {Modak}, \citenamefont {Srivastava},\ and\
  \citenamefont {Valle}}]{1705.00915}%
  \BibitemOpen
  \bibfield  {author} {\bibinfo {author} {\bibfnamefont {C.}~\bibnamefont
  {Bonilla}}, \bibinfo {author} {\bibfnamefont {T.}~\bibnamefont {Modak}},
  \bibinfo {author} {\bibfnamefont {R.}~\bibnamefont {Srivastava}}, \ and\
  \bibinfo {author} {\bibfnamefont {J.~W.~F.}\ \bibnamefont {Valle}},\
  }\href@noop {} {\  (\bibinfo {year} {2017})},\ \Eprint
  {http://arxiv.org/abs/1705.00915} {arXiv:1705.00915 [hep-ph]} \BibitemShut
  {NoStop}%
\bibitem [{\citenamefont {Babu}\ \emph {et~al.}(2017)\citenamefont {Babu},
  \citenamefont {Friedland}, \citenamefont {Machado},\ and\ \citenamefont
  {Mocioiu}}]{1705.01822}%
  \BibitemOpen
  \bibfield  {author} {\bibinfo {author} {\bibfnamefont {K.~S.}\ \bibnamefont
  {Babu}}, \bibinfo {author} {\bibfnamefont {A.}~\bibnamefont {Friedland}},
  \bibinfo {author} {\bibfnamefont {P.~A.~N.}\ \bibnamefont {Machado}}, \ and\
  \bibinfo {author} {\bibfnamefont {I.}~\bibnamefont {Mocioiu}},\ }\href@noop
  {} {\  (\bibinfo {year} {2017})},\ \Eprint {http://arxiv.org/abs/1705.01822}
  {arXiv:1705.01822 [hep-ph]} \BibitemShut {NoStop}%
\bibitem [{\citenamefont {Alonso}\ \emph
  {et~al.}(2017{\natexlab{b}})\citenamefont {Alonso}, \citenamefont {Cox},
  \citenamefont {Han},\ and\ \citenamefont {Yanagida}}]{1705.03858}%
  \BibitemOpen
  \bibfield  {author} {\bibinfo {author} {\bibfnamefont {R.}~\bibnamefont
  {Alonso}}, \bibinfo {author} {\bibfnamefont {P.}~\bibnamefont {Cox}},
  \bibinfo {author} {\bibfnamefont {C.}~\bibnamefont {Han}}, \ and\ \bibinfo
  {author} {\bibfnamefont {T.~T.}\ \bibnamefont {Yanagida}},\ }\href {\doibase
  10.1016/j.physletb.2017.10.027} {\bibfield  {journal} {\bibinfo  {journal}
  {Phys. Lett.}\ }\textbf {\bibinfo {volume} {B774}},\ \bibinfo {pages} {643}
  (\bibinfo {year} {2017}{\natexlab{b}})},\ \Eprint
  {http://arxiv.org/abs/1705.03858} {arXiv:1705.03858 [hep-ph]} \BibitemShut
  {NoStop}%
\bibitem [{\citenamefont {Okada}\ and\ \citenamefont {Seto}(2010)}]{1002.2525}%
  \BibitemOpen
  \bibfield  {author} {\bibinfo {author} {\bibfnamefont {N.}~\bibnamefont
  {Okada}}\ and\ \bibinfo {author} {\bibfnamefont {O.}~\bibnamefont {Seto}},\
  }\href {\doibase 10.1103/PhysRevD.82.023507} {\bibfield  {journal} {\bibinfo
  {journal} {Phys. Rev.}\ }\textbf {\bibinfo {volume} {D82}},\ \bibinfo {pages}
  {023507} (\bibinfo {year} {2010})},\ \Eprint {http://arxiv.org/abs/1002.2525}
  {arXiv:1002.2525 [hep-ph]} \BibitemShut {NoStop}%
\bibitem [{\citenamefont {Okada}\ and\ \citenamefont
  {Okada}(2016)}]{1601.07526}%
  \BibitemOpen
  \bibfield  {author} {\bibinfo {author} {\bibfnamefont {N.}~\bibnamefont
  {Okada}}\ and\ \bibinfo {author} {\bibfnamefont {S.}~\bibnamefont {Okada}},\
  }\href {\doibase 10.1103/PhysRevD.93.075003} {\bibfield  {journal} {\bibinfo
  {journal} {Phys. Rev.}\ }\textbf {\bibinfo {volume} {D93}},\ \bibinfo {pages}
  {075003} (\bibinfo {year} {2016})},\ \Eprint
  {http://arxiv.org/abs/1601.07526} {arXiv:1601.07526 [hep-ph]} \BibitemShut
  {NoStop}%
\bibitem [{\citenamefont {Kaneta}\ \emph {et~al.}(2017)\citenamefont {Kaneta},
  \citenamefont {Kang},\ and\ \citenamefont {Lee}}]{1606.09317}%
  \BibitemOpen
  \bibfield  {author} {\bibinfo {author} {\bibfnamefont {K.}~\bibnamefont
  {Kaneta}}, \bibinfo {author} {\bibfnamefont {Z.}~\bibnamefont {Kang}}, \ and\
  \bibinfo {author} {\bibfnamefont {H.-S.}\ \bibnamefont {Lee}},\ }\href
  {\doibase 10.1007/JHEP02(2017)031} {\bibfield  {journal} {\bibinfo  {journal}
  {JHEP}\ }\textbf {\bibinfo {volume} {02}},\ \bibinfo {pages} {031} (\bibinfo
  {year} {2017})},\ \Eprint {http://arxiv.org/abs/1606.09317} {arXiv:1606.09317
  [hep-ph]} \BibitemShut {NoStop}%
\bibitem [{\citenamefont {Okada}\ and\ \citenamefont
  {Okada}(2017)}]{1611.02672}%
  \BibitemOpen
  \bibfield  {author} {\bibinfo {author} {\bibfnamefont {N.}~\bibnamefont
  {Okada}}\ and\ \bibinfo {author} {\bibfnamefont {S.}~\bibnamefont {Okada}},\
  }\href {\doibase 10.1103/PhysRevD.95.035025} {\bibfield  {journal} {\bibinfo
  {journal} {Phys. Rev.}\ }\textbf {\bibinfo {volume} {D95}},\ \bibinfo {pages}
  {035025} (\bibinfo {year} {2017})},\ \Eprint
  {http://arxiv.org/abs/1611.02672} {arXiv:1611.02672 [hep-ph]} \BibitemShut
  {NoStop}%
\bibitem [{\citenamefont {Duerr}\ \emph {et~al.}(2016)\citenamefont {Duerr},
  \citenamefont {Kahlhoefer}, \citenamefont {Schmidt-Hoberg}, \citenamefont
  {Schwetz},\ and\ \citenamefont {Vogl}}]{1606.07609}%
  \BibitemOpen
  \bibfield  {author} {\bibinfo {author} {\bibfnamefont {M.}~\bibnamefont
  {Duerr}}, \bibinfo {author} {\bibfnamefont {F.}~\bibnamefont {Kahlhoefer}},
  \bibinfo {author} {\bibfnamefont {K.}~\bibnamefont {Schmidt-Hoberg}},
  \bibinfo {author} {\bibfnamefont {T.}~\bibnamefont {Schwetz}}, \ and\
  \bibinfo {author} {\bibfnamefont {S.}~\bibnamefont {Vogl}},\ }\href {\doibase
  10.1007/JHEP09(2016)042} {\bibfield  {journal} {\bibinfo  {journal} {JHEP}\
  }\textbf {\bibinfo {volume} {09}},\ \bibinfo {pages} {042} (\bibinfo {year}
  {2016})},\ \Eprint {http://arxiv.org/abs/1606.07609} {arXiv:1606.07609
  [hep-ph]} \BibitemShut {NoStop}%
\bibitem [{\citenamefont {Bell}\ \emph {et~al.}(2016)\citenamefont {Bell},
  \citenamefont {Cai},\ and\ \citenamefont {Leane}}]{1605.09382}%
  \BibitemOpen
  \bibfield  {author} {\bibinfo {author} {\bibfnamefont {N.~F.}\ \bibnamefont
  {Bell}}, \bibinfo {author} {\bibfnamefont {Y.}~\bibnamefont {Cai}}, \ and\
  \bibinfo {author} {\bibfnamefont {R.~K.}\ \bibnamefont {Leane}},\ }\href
  {\doibase 10.1088/1475-7516/2016/08/001} {\bibfield  {journal} {\bibinfo
  {journal} {JCAP}\ }\textbf {\bibinfo {volume} {1608}},\ \bibinfo {pages}
  {001} (\bibinfo {year} {2016})},\ \Eprint {http://arxiv.org/abs/1605.09382}
  {arXiv:1605.09382 [hep-ph]} \BibitemShut {NoStop}%
\bibitem [{\citenamefont {Babu}\ \emph {et~al.}(2008)\citenamefont {Babu},
  \citenamefont {Barr},\ and\ \citenamefont {Gogoladze}}]{0709.3491}%
  \BibitemOpen
  \bibfield  {author} {\bibinfo {author} {\bibfnamefont {K.~S.}\ \bibnamefont
  {Babu}}, \bibinfo {author} {\bibfnamefont {S.~M.}\ \bibnamefont {Barr}}, \
  and\ \bibinfo {author} {\bibfnamefont {I.}~\bibnamefont {Gogoladze}},\ }\href
  {\doibase 10.1016/j.physletb.2008.01.057} {\bibfield  {journal} {\bibinfo
  {journal} {Phys. Lett.}\ }\textbf {\bibinfo {volume} {B661}},\ \bibinfo
  {pages} {124} (\bibinfo {year} {2008})},\ \Eprint
  {http://arxiv.org/abs/0709.3491} {arXiv:0709.3491 [hep-ph]} \BibitemShut
  {NoStop}%
\bibitem [{\citenamefont {Aaij}\ \emph {et~al.}(2014)\citenamefont {Aaij} \emph
  {et~al.}}]{1406.6482}%
  \BibitemOpen
  \bibfield  {author} {\bibinfo {author} {\bibfnamefont {R.}~\bibnamefont
  {Aaij}} \emph {et~al.} (\bibinfo {collaboration} {LHCb}),\ }\href {\doibase
  10.1103/PhysRevLett.113.151601} {\bibfield  {journal} {\bibinfo  {journal}
  {Phys. Rev. Lett.}\ }\textbf {\bibinfo {volume} {113}},\ \bibinfo {pages}
  {151601} (\bibinfo {year} {2014})},\ \Eprint {http://arxiv.org/abs/1406.6482}
  {arXiv:1406.6482 [hep-ex]} \BibitemShut {NoStop}%
\bibitem [{\citenamefont {Aaij}\ \emph {et~al.}(2017)\citenamefont {Aaij} \emph
  {et~al.}}]{1705.05802}%
  \BibitemOpen
  \bibfield  {author} {\bibinfo {author} {\bibfnamefont {R.}~\bibnamefont
  {Aaij}} \emph {et~al.} (\bibinfo {collaboration} {LHCb}),\ }\href {\doibase
  10.1007/JHEP08(2017)055} {\bibfield  {journal} {\bibinfo  {journal} {JHEP}\
  }\textbf {\bibinfo {volume} {08}},\ \bibinfo {pages} {055} (\bibinfo {year}
  {2017})},\ \Eprint {http://arxiv.org/abs/1705.05802} {arXiv:1705.05802
  [hep-ex]} \BibitemShut {NoStop}%
\bibitem [{\citenamefont {Bélanger}\ \emph {et~al.}(2015)\citenamefont
  {Bélanger}, \citenamefont {Boudjema}, \citenamefont {Pukhov},\ and\
  \citenamefont {Semenov}}]{1407.6129}%
  \BibitemOpen
  \bibfield  {author} {\bibinfo {author} {\bibfnamefont {G.}~\bibnamefont
  {Bélanger}}, \bibinfo {author} {\bibfnamefont {F.}~\bibnamefont {Boudjema}},
  \bibinfo {author} {\bibfnamefont {A.}~\bibnamefont {Pukhov}}, \ and\ \bibinfo
  {author} {\bibfnamefont {A.}~\bibnamefont {Semenov}},\ }\href {\doibase
  10.1016/j.cpc.2015.03.003} {\bibfield  {journal} {\bibinfo  {journal}
  {Comput. Phys. Commun.}\ }\textbf {\bibinfo {volume} {192}},\ \bibinfo
  {pages} {322} (\bibinfo {year} {2015})},\ \Eprint
  {http://arxiv.org/abs/1407.6129} {arXiv:1407.6129 [hep-ph]} \BibitemShut
  {NoStop}%
\bibitem [{\citenamefont {Lee}\ \emph {et~al.}(1977{\natexlab{a}})\citenamefont
  {Lee}, \citenamefont {Quigg},\ and\ \citenamefont {Thacker}}]{Lee:1977yc}%
  \BibitemOpen
  \bibfield  {author} {\bibinfo {author} {\bibfnamefont {B.~W.}\ \bibnamefont
  {Lee}}, \bibinfo {author} {\bibfnamefont {C.}~\bibnamefont {Quigg}}, \ and\
  \bibinfo {author} {\bibfnamefont {H.~B.}\ \bibnamefont {Thacker}},\ }\href
  {\doibase 10.1103/PhysRevLett.38.883} {\bibfield  {journal} {\bibinfo
  {journal} {Phys. Rev. Lett.}\ }\textbf {\bibinfo {volume} {38}},\ \bibinfo
  {pages} {883} (\bibinfo {year} {1977}{\natexlab{a}})}\BibitemShut {NoStop}%
\bibitem [{\citenamefont {Lee}\ \emph {et~al.}(1977{\natexlab{b}})\citenamefont
  {Lee}, \citenamefont {Quigg},\ and\ \citenamefont {Thacker}}]{Lee:1977eg}%
  \BibitemOpen
  \bibfield  {author} {\bibinfo {author} {\bibfnamefont {B.~W.}\ \bibnamefont
  {Lee}}, \bibinfo {author} {\bibfnamefont {C.}~\bibnamefont {Quigg}}, \ and\
  \bibinfo {author} {\bibfnamefont {H.~B.}\ \bibnamefont {Thacker}},\ }\href
  {\doibase 10.1103/PhysRevD.16.1519} {\bibfield  {journal} {\bibinfo
  {journal} {Phys. Rev.}\ }\textbf {\bibinfo {volume} {D16}},\ \bibinfo {pages}
  {1519} (\bibinfo {year} {1977}{\natexlab{b}})}\BibitemShut {NoStop}%
\bibitem [{\citenamefont {Chanowitz}\ \emph {et~al.}(1979)\citenamefont
  {Chanowitz}, \citenamefont {Furman},\ and\ \citenamefont
  {Hinchliffe}}]{Chanowitz:1978mv}%
  \BibitemOpen
  \bibfield  {author} {\bibinfo {author} {\bibfnamefont {M.~S.}\ \bibnamefont
  {Chanowitz}}, \bibinfo {author} {\bibfnamefont {M.~A.}\ \bibnamefont
  {Furman}}, \ and\ \bibinfo {author} {\bibfnamefont {I.}~\bibnamefont
  {Hinchliffe}},\ }\href {\doibase 10.1016/0550-3213(79)90606-0} {\bibfield
  {journal} {\bibinfo  {journal} {Nucl. Phys.}\ }\textbf {\bibinfo {volume}
  {B153}},\ \bibinfo {pages} {402} (\bibinfo {year} {1979})}\BibitemShut
  {NoStop}%
\bibitem [{\citenamefont {El~Hedri}\ \emph {et~al.}(2017)\citenamefont
  {El~Hedri}, \citenamefont {Shepherd},\ and\ \citenamefont
  {Walker}}]{1412.5660}%
  \BibitemOpen
  \bibfield  {author} {\bibinfo {author} {\bibfnamefont {S.}~\bibnamefont
  {El~Hedri}}, \bibinfo {author} {\bibfnamefont {W.}~\bibnamefont {Shepherd}},
  \ and\ \bibinfo {author} {\bibfnamefont {D.~G.~E.}\ \bibnamefont {Walker}},\
  }\href {\doibase 10.1016/j.dark.2017.09.006} {\bibfield  {journal} {\bibinfo
  {journal} {Phys. Dark Univ.}\ }\textbf {\bibinfo {volume} {18}},\ \bibinfo
  {pages} {127} (\bibinfo {year} {2017})},\ \Eprint
  {http://arxiv.org/abs/1412.5660} {arXiv:1412.5660 [hep-ph]} \BibitemShut
  {NoStop}%
\bibitem [{\citenamefont {Kahlhoefer}\ \emph {et~al.}(2016)\citenamefont
  {Kahlhoefer}, \citenamefont {Schmidt-Hoberg}, \citenamefont {Schwetz},\ and\
  \citenamefont {Vogl}}]{1510.02110}%
  \BibitemOpen
  \bibfield  {author} {\bibinfo {author} {\bibfnamefont {F.}~\bibnamefont
  {Kahlhoefer}}, \bibinfo {author} {\bibfnamefont {K.}~\bibnamefont
  {Schmidt-Hoberg}}, \bibinfo {author} {\bibfnamefont {T.}~\bibnamefont
  {Schwetz}}, \ and\ \bibinfo {author} {\bibfnamefont {S.}~\bibnamefont
  {Vogl}},\ }\href {\doibase 10.1007/JHEP02(2016)016} {\bibfield  {journal}
  {\bibinfo  {journal} {JHEP}\ }\textbf {\bibinfo {volume} {02}},\ \bibinfo
  {pages} {016} (\bibinfo {year} {2016})},\ \Eprint
  {http://arxiv.org/abs/1510.02110} {arXiv:1510.02110 [hep-ph]} \BibitemShut
  {NoStop}%
\bibitem [{\citenamefont {Hook}\ \emph {et~al.}(2011)\citenamefont {Hook},
  \citenamefont {Izaguirre},\ and\ \citenamefont {Wacker}}]{1006.0973}%
  \BibitemOpen
  \bibfield  {author} {\bibinfo {author} {\bibfnamefont {A.}~\bibnamefont
  {Hook}}, \bibinfo {author} {\bibfnamefont {E.}~\bibnamefont {Izaguirre}}, \
  and\ \bibinfo {author} {\bibfnamefont {J.~G.}\ \bibnamefont {Wacker}},\
  }\href {\doibase 10.1155/2011/859762} {\bibfield  {journal} {\bibinfo
  {journal} {Adv. High Energy Phys.}\ }\textbf {\bibinfo {volume} {2011}},\
  \bibinfo {pages} {859762} (\bibinfo {year} {2011})},\ \Eprint
  {http://arxiv.org/abs/1006.0973} {arXiv:1006.0973 [hep-ph]} \BibitemShut
  {NoStop}%
\bibitem [{\citenamefont {Aad}\ \emph {et~al.}(2016)\citenamefont {Aad} \emph
  {et~al.}}]{1606.02266}%
  \BibitemOpen
  \bibfield  {author} {\bibinfo {author} {\bibfnamefont {G.}~\bibnamefont
  {Aad}} \emph {et~al.} (\bibinfo {collaboration} {ATLAS, CMS}),\ }\href
  {\doibase 10.1007/JHEP08(2016)045} {\bibfield  {journal} {\bibinfo  {journal}
  {JHEP}\ }\textbf {\bibinfo {volume} {08}},\ \bibinfo {pages} {045} (\bibinfo
  {year} {2016})},\ \Eprint {http://arxiv.org/abs/1606.02266} {arXiv:1606.02266
  [hep-ex]} \BibitemShut {NoStop}%
\bibitem [{\citenamefont {Aad}\ \emph {et~al.}(2015)\citenamefont {Aad} \emph
  {et~al.}}]{1509.00672}%
  \BibitemOpen
  \bibfield  {author} {\bibinfo {author} {\bibfnamefont {G.}~\bibnamefont
  {Aad}} \emph {et~al.} (\bibinfo {collaboration} {ATLAS}),\ }\href {\doibase
  10.1007/JHEP11(2015)206} {\bibfield  {journal} {\bibinfo  {journal} {JHEP}\
  }\textbf {\bibinfo {volume} {11}},\ \bibinfo {pages} {206} (\bibinfo {year}
  {2015})},\ \Eprint {http://arxiv.org/abs/1509.00672} {arXiv:1509.00672
  [hep-ex]} \BibitemShut {NoStop}%
\bibitem [{\citenamefont {Khachatryan}\ \emph {et~al.}(2017)\citenamefont
  {Khachatryan} \emph {et~al.}}]{1610.09218}%
  \BibitemOpen
  \bibfield  {author} {\bibinfo {author} {\bibfnamefont {V.}~\bibnamefont
  {Khachatryan}} \emph {et~al.} (\bibinfo {collaboration} {CMS}),\ }\href
  {\doibase 10.1007/JHEP02(2017)135} {\bibfield  {journal} {\bibinfo  {journal}
  {JHEP}\ }\textbf {\bibinfo {volume} {02}},\ \bibinfo {pages} {135} (\bibinfo
  {year} {2017})},\ \Eprint {http://arxiv.org/abs/1610.09218} {arXiv:1610.09218
  [hep-ex]} \BibitemShut {NoStop}%
\bibitem [{\citenamefont {del Amo~Sanchez}\ \emph {et~al.}(2010)\citenamefont
  {del Amo~Sanchez} \emph {et~al.}}]{1002.4358}%
  \BibitemOpen
  \bibfield  {author} {\bibinfo {author} {\bibfnamefont {P.}~\bibnamefont {del
  Amo~Sanchez}} \emph {et~al.} (\bibinfo {collaboration} {BaBar}),\ }\href
  {\doibase 10.1103/PhysRevLett.104.191801} {\bibfield  {journal} {\bibinfo
  {journal} {Phys. Rev. Lett.}\ }\textbf {\bibinfo {volume} {104}},\ \bibinfo
  {pages} {191801} (\bibinfo {year} {2010})},\ \Eprint
  {http://arxiv.org/abs/1002.4358} {arXiv:1002.4358 [hep-ex]} \BibitemShut
  {NoStop}%
\bibitem [{\citenamefont {D'Eramo}\ \emph {et~al.}(2016)\citenamefont
  {D'Eramo}, \citenamefont {Kavanagh},\ and\ \citenamefont
  {Panci}}]{1605.04917}%
  \BibitemOpen
  \bibfield  {author} {\bibinfo {author} {\bibfnamefont {F.}~\bibnamefont
  {D'Eramo}}, \bibinfo {author} {\bibfnamefont {B.~J.}\ \bibnamefont
  {Kavanagh}}, \ and\ \bibinfo {author} {\bibfnamefont {P.}~\bibnamefont
  {Panci}},\ }\href {\doibase 10.1007/JHEP08(2016)111} {\bibfield  {journal}
  {\bibinfo  {journal} {JHEP}\ }\textbf {\bibinfo {volume} {08}},\ \bibinfo
  {pages} {111} (\bibinfo {year} {2016})},\ \Eprint
  {http://arxiv.org/abs/1605.04917} {arXiv:1605.04917 [hep-ph]} \BibitemShut
  {NoStop}%
\bibitem [{\citenamefont {Aprile}\ \emph {et~al.}(2017)\citenamefont {Aprile}
  \emph {et~al.}}]{1705.06655}%
  \BibitemOpen
  \bibfield  {author} {\bibinfo {author} {\bibfnamefont {E.}~\bibnamefont
  {Aprile}} \emph {et~al.} (\bibinfo {collaboration} {XENON}),\ }\href
  {\doibase 10.1103/PhysRevLett.119.181301} {\bibfield  {journal} {\bibinfo
  {journal} {Phys. Rev. Lett.}\ }\textbf {\bibinfo {volume} {119}},\ \bibinfo
  {pages} {181301} (\bibinfo {year} {2017})},\ \Eprint
  {http://arxiv.org/abs/1705.06655} {arXiv:1705.06655 [astro-ph.CO]}
  \BibitemShut {NoStop}%
\bibitem [{\citenamefont {Albert}\ \emph {et~al.}(2017)\citenamefont {Albert}
  \emph {et~al.}}]{1611.03184}%
  \BibitemOpen
  \bibfield  {author} {\bibinfo {author} {\bibfnamefont {A.}~\bibnamefont
  {Albert}} \emph {et~al.} (\bibinfo {collaboration} {DES, Fermi-LAT}),\ }\href
  {\doibase 10.3847/1538-4357/834/2/110} {\bibfield  {journal} {\bibinfo
  {journal} {Astrophys. J.}\ }\textbf {\bibinfo {volume} {834}},\ \bibinfo
  {pages} {110} (\bibinfo {year} {2017})},\ \Eprint
  {http://arxiv.org/abs/1611.03184} {arXiv:1611.03184 [astro-ph.HE]}
  \BibitemShut {NoStop}%
\bibitem [{\citenamefont {Sjostrand}\ \emph {et~al.}(2006)\citenamefont
  {Sjostrand}, \citenamefont {Mrenna},\ and\ \citenamefont
  {Skands}}]{hep-ph/0603175}%
  \BibitemOpen
  \bibfield  {author} {\bibinfo {author} {\bibfnamefont {T.}~\bibnamefont
  {Sjostrand}}, \bibinfo {author} {\bibfnamefont {S.}~\bibnamefont {Mrenna}}, \
  and\ \bibinfo {author} {\bibfnamefont {P.~Z.}\ \bibnamefont {Skands}},\
  }\href {\doibase 10.1088/1126-6708/2006/05/026} {\bibfield  {journal}
  {\bibinfo  {journal} {JHEP}\ }\textbf {\bibinfo {volume} {05}},\ \bibinfo
  {pages} {026} (\bibinfo {year} {2006})},\ \Eprint
  {http://arxiv.org/abs/hep-ph/0603175} {arXiv:hep-ph/0603175 [hep-ph]}
  \BibitemShut {NoStop}%
\bibitem [{\citenamefont {Ackermann}\ \emph {et~al.}(2015)\citenamefont
  {Ackermann} \emph {et~al.}}]{1503.02641}%
  \BibitemOpen
  \bibfield  {author} {\bibinfo {author} {\bibfnamefont {M.}~\bibnamefont
  {Ackermann}} \emph {et~al.} (\bibinfo {collaboration} {Fermi-LAT}),\ }\href
  {\doibase 10.1103/PhysRevLett.115.231301} {\bibfield  {journal} {\bibinfo
  {journal} {Phys. Rev. Lett.}\ }\textbf {\bibinfo {volume} {115}},\ \bibinfo
  {pages} {231301} (\bibinfo {year} {2015})},\ \Eprint
  {http://arxiv.org/abs/1503.02641} {arXiv:1503.02641 [astro-ph.HE]}
  \BibitemShut {NoStop}%
\bibitem [{\citenamefont {Alwall}\ \emph {et~al.}(2014)\citenamefont {Alwall},
  \citenamefont {Frederix}, \citenamefont {Frixione}, \citenamefont {Hirschi},
  \citenamefont {Maltoni}, \citenamefont {Mattelaer}, \citenamefont {Shao},
  \citenamefont {Stelzer}, \citenamefont {Torrielli},\ and\ \citenamefont
  {Zaro}}]{1405.0301}%
  \BibitemOpen
  \bibfield  {author} {\bibinfo {author} {\bibfnamefont {J.}~\bibnamefont
  {Alwall}}, \bibinfo {author} {\bibfnamefont {R.}~\bibnamefont {Frederix}},
  \bibinfo {author} {\bibfnamefont {S.}~\bibnamefont {Frixione}}, \bibinfo
  {author} {\bibfnamefont {V.}~\bibnamefont {Hirschi}}, \bibinfo {author}
  {\bibfnamefont {F.}~\bibnamefont {Maltoni}}, \bibinfo {author} {\bibfnamefont
  {O.}~\bibnamefont {Mattelaer}}, \bibinfo {author} {\bibfnamefont {H.~S.}\
  \bibnamefont {Shao}}, \bibinfo {author} {\bibfnamefont {T.}~\bibnamefont
  {Stelzer}}, \bibinfo {author} {\bibfnamefont {P.}~\bibnamefont {Torrielli}},
  \ and\ \bibinfo {author} {\bibfnamefont {M.}~\bibnamefont {Zaro}},\ }\href
  {\doibase 10.1007/JHEP07(2014)079} {\bibfield  {journal} {\bibinfo  {journal}
  {JHEP}\ }\textbf {\bibinfo {volume} {07}},\ \bibinfo {pages} {079} (\bibinfo
  {year} {2014})},\ \Eprint {http://arxiv.org/abs/1405.0301} {arXiv:1405.0301
  [hep-ph]} \BibitemShut {NoStop}%
\bibitem [{\citenamefont {Sirunyan}\ \emph {et~al.}(2017)\citenamefont
  {Sirunyan} \emph {et~al.}}]{1703.01651}%
  \BibitemOpen
  \bibfield  {author} {\bibinfo {author} {\bibfnamefont {A.~M.}\ \bibnamefont
  {Sirunyan}} \emph {et~al.} (\bibinfo {collaboration} {CMS}),\ }\href
  {\doibase 10.1007/JHEP07(2017)014} {\bibfield  {journal} {\bibinfo  {journal}
  {JHEP}\ }\textbf {\bibinfo {volume} {07}},\ \bibinfo {pages} {014} (\bibinfo
  {year} {2017})},\ \Eprint {http://arxiv.org/abs/1703.01651} {arXiv:1703.01651
  [hep-ex]} \BibitemShut {NoStop}%
\bibitem [{ATL(2017)}]{ATLAS-CONF-2017-060}%
  \BibitemOpen
  ATLAS Collaboration,\ \bibinfo {type} {Tech. Rep.}\ \bibinfo {number}
  {ATLAS-CONF-2017-060}\ (\bibinfo  {institution} {CERN},\ \bibinfo {address}
  {Geneva},\ \bibinfo {year} {2017})\BibitemShut {NoStop}%
\bibitem [{\citenamefont {Aaboud}\ \emph {et~al.}(2017)\citenamefont {Aaboud}
  \emph {et~al.}}]{1709.07242}%
  \BibitemOpen
  \bibfield  {author} {\bibinfo {author} {\bibfnamefont {M.}~\bibnamefont
  {Aaboud}} \emph {et~al.} (\bibinfo {collaboration} {ATLAS}),\ }\href@noop {}
  {\  (\bibinfo {year} {2017})},\ \Eprint {http://arxiv.org/abs/1709.07242}
  {arXiv:1709.07242 [hep-ex]} \BibitemShut {NoStop}%
\bibitem [{\citenamefont {Carr}\ \emph {et~al.}(2016)\citenamefont {Carr} \emph
  {et~al.}}]{1508.06128}%
  \BibitemOpen
  \bibfield  {author} {\bibinfo {author} {\bibfnamefont {J.}~\bibnamefont
  {Carr}} \emph {et~al.} (\bibinfo {collaboration} {CTA}),\ }\bibfield
  {booktitle} {\emph {\bibinfo {booktitle} {{Proceedings, 34th International
  Cosmic Ray Conference (ICRC 2015): The Hague, The Netherlands, July 30-August
  6, 2015}}},\ }\href@noop {} {\bibfield  {journal} {\bibinfo  {journal} {PoS}\
  }\textbf {\bibinfo {volume} {ICRC2015}},\ \bibinfo {pages} {1203} (\bibinfo
  {year} {2016})},\ \bibinfo {note} {[34,1203(2015)]},\ \Eprint
  {http://arxiv.org/abs/1508.06128} {arXiv:1508.06128 [astro-ph.HE]}
  \BibitemShut {NoStop}%
\bibitem [{\citenamefont {Liu}\ \emph {et~al.}(2017)\citenamefont {Liu},
  \citenamefont {Wang},\ and\ \citenamefont {Yu}}]{1704.00730}%
  \BibitemOpen
  \bibfield  {author} {\bibinfo {author} {\bibfnamefont {J.}~\bibnamefont
  {Liu}}, \bibinfo {author} {\bibfnamefont {X.-P.}\ \bibnamefont {Wang}}, \
  and\ \bibinfo {author} {\bibfnamefont {F.}~\bibnamefont {Yu}},\ }\href
  {\doibase 10.1007/JHEP06(2017)077} {\bibfield  {journal} {\bibinfo  {journal}
  {JHEP}\ }\textbf {\bibinfo {volume} {06}},\ \bibinfo {pages} {077} (\bibinfo
  {year} {2017})},\ \Eprint {http://arxiv.org/abs/1704.00730} {arXiv:1704.00730
  [hep-ph]} \BibitemShut {NoStop}%
\end{thebibliography}%

\end{document}